\documentclass[twocolumn,showpacs,amsmatsh,amssymb,aps,prc,groupedaddress,superscriptaddress,nofootinbib]{revtex4-1}

\usepackage[caption=false]{subfig}
\usepackage{tabularx}
\usepackage{graphicx}
\graphicspath{{./figs/}}
\usepackage{amsmath}
\usepackage{todonotes}
\usepackage{hyperref}
\allowdisplaybreaks

\begin{document}


\title{Influence of short-range correlations in neutrino-nucleus scattering}


\author{T.~Van Cuyck}
\email{Tom.VanCuyck@UGent.be}
\affiliation{Department of Physics and Astronomy,\\
Ghent University,\\ Proeftuinstraat 86,\\ B-9000 Gent, Belgium}
\author{N.~Jachowicz}
\email{Natalie.Jachowicz@UGent.be}
\affiliation{Department of Physics and Astronomy,\\
Ghent University,\\ Proeftuinstraat 86,\\ B-9000 Gent, Belgium}
\author{R.~Gonz\'{a}lez-Jim\'{e}nez}
\affiliation{Department of Physics and Astronomy,\\
Ghent University,\\ Proeftuinstraat 86,\\ B-9000 Gent, Belgium}
\author{M.~Martini}
\affiliation{Department of Physics and Astronomy,\\
Ghent University,\\ Proeftuinstraat 86,\\ B-9000 Gent, Belgium}
\affiliation{ESNT, CEA-Saclay, IRFU,\\
Service de Physique Nucl\'eaire, \\ F-91191 Gif-sur-Yvette Cedex, France}
\author{V.~Pandey}
\affiliation{Department of Physics and Astronomy,\\
Ghent University,\\ Proeftuinstraat 86,\\ B-9000 Gent, Belgium}
\author{J.~Ryckebusch}
\affiliation{Department of Physics and Astronomy,\\
Ghent University,\\ Proeftuinstraat 86,\\ B-9000 Gent, Belgium}
\author{N.~Van Dessel}
\affiliation{Department of Physics and Astronomy,\\
Ghent University,\\ Proeftuinstraat 86,\\ B-9000 Gent, Belgium}



\date{\today}

\begin{abstract}
  \begin{description}  
    \item[Background] Nuclear short-range correlations (SRCs) are corrections to mean-field wave functions connected with the short-distance behavior of the nucleon-nucleon interaction. These SRCs provide corrections to lepton-nucleus cross sections as computed in the impulse approximation (IA).
    \item[Purpose] We want to investigate the influence of SRCs on the one-nucleon ($1N$) and two-nucleon ($2N$) knockout channel for muon-neutrino induced processes on a $^{12}$C target at energies relevant for contemporary measurements.
    \item[Method] The model adopted in this work, corrects the impulse approximation for SRCs by shifting the complexity induced by the SRCs from the wave functions to the operators. Due to the local character of the SRCs, it is argued that the expansion of these operators can be truncated at a low order.
    \item[Results] The model is compared with electron-scattering data, and two-particle two-hole responses are presented for neutrino scattering. The contributions from the vector and axial-vector parts of the nuclear current as well as the central, tensor and spin-isospin part of the SRCs are studied.
    \item[Conclusions] Nuclear SRCs affect the $1N$ knockout channel and give rise to $2N$ knockout. The exclusive neutrino-induced $2N$ knockout cross section of SRC pairs is shown and the $2N$ knockout contribution to the QE signal is calculated. The strength occurs as a broad background which extends into the dip region.
  \end{description}
\end{abstract}


\maketitle

\section{Introduction}
One of the major issues in neutrino-scattering studies is the contribution of two-body currents to the measured quasielastic-like neutrino-nucleus ($\nu A$) cross section. A thorough knowledge of this contribution is necessary for a rigorous description of $\nu A$ cross sections at intermediate (0.1 - 2 GeV) energies. A genuine quasielastic (QE) calculation, where the $W$ boson interacts with a single nucleon which leads to a one-particle one-hole (1p1h) excitation, does not accurately describe recent measurements of 
neutrino ($\nu$) and antineutrino ($\overline{\nu}$) cross sections \cite{AguilarArevalo:2010zc,AguilarArevalo:2013dva,Fiorentini:2013ezn,Fields:2013zhk,Abe:2015oar,Abe:2014iza,Abe:2016tmq}. 
Since typical $\nu_\mu A$ measurements do not uniquely determine the nuclear final state as only the energy and momentum of the muon are measured, 
the absorption of the $W$ boson by a single nucleon is only one of the many possible interaction mechanisms.
In addition  one must consider coupling to nucleons belonging to short-range correlation (SRC) pairs and to two-body currents arising from meson-exchange currents (MECs). 
This leads to multinucleon excitations, of which the two-particle two-hole (2p2h) ones constitute the leading order.
Several theoretical approaches have analyzed the role of multinucleon excitations in the $\nu A$ cross sections by comparing their results with experimental data \cite{Kolbe:1994xb,Kim:1994zea,Martini:2009uj,Martini:2010ex,Amaro:2010sd,Nieves:2011pp,Bodek:2011ps,Martini:2011wp,Nieves:2011yp,Amaro:2011aa,Lalakulich:2012ac,Nieves:2013fr,Martini:2013sha,Gran:2013kda,Martini:2014dqa,Megias:2014qva,Ericson:2015cva,Ivanov:2015aya,Martini:2016eec}. A
complete theoretical model should in principle include short-range and long-range nuclear correlations, MEC and final-state interactions (FSIs). In this work, we focus on the influence of nuclear SRCs on inclusive QE cross sections.
Different models which account for multinucleon effects in $\nu A$ and $\overline{\nu} A$ reactions have been developed \cite{Alvarez-Ruso:2014bla}. These are the microscopic models of Martini \textit{et~al.}~\cite{Martini:2009uj} and Nieves \textit{et~al.}~\cite{Nieves:2011pp} and the superscaling approach (SuSA)~\cite{Amaro:2010sd}. 
Summarizing, the models by Martini \textit{et~al.}~and Nieves \textit{et~al.}~take nuclear finite-size effects into account via a local density approximation and a semi-classical expansion of the response function, but ignore the shell structure which is taken into account in Refs.~\cite{Pandey:2013cca,Pandey:2014tza}. Long-range RPA correlations are taken into account in Refs.~\cite{Martini:2009uj,Nieves:2011pp,Pandey:2013cca,Pandey:2014tza}. In the 2p2h sector, the two
models are based on the Fermi gas, which is the simplest independent-particle model (IPM). Both approaches consider two-body MEC contributions. The nucleon-nucleon SRCs are included by considering an additional two-body current, the correlation current. With the introduction of the correlation contributions, the interference between correlations and MECs naturally appears. 
In the SuSA approach, a superscaling analysis of electron scattering results is used to predict $\nu A$ cross sections~\cite{Amaro:2004bs}. The effects of SRCs and MECs in the 1p1h sector are effectively included via the phenomenological superscaling function. In~\cite{Megias:2014qva}, the SuSA model is combined with MECs in the 2p2h sector, by using a parameterization of the microscopic calculations by De Pace \textit{et al.}~\cite{DePace:2003xu}. The correlations and correlations-MEC interference terms are absent
in the 2p2h channel. 
A relativistic Fermi gas (RFG)-based model that accounts for MECs, correlations and interference in the 1p1h and 2p2h sector for electron-nucleus ($eA$) scattering has been developed by Amaro \textit{et~al.}~\cite{Amaro:2002mj,Amaro:2010iu}, which has recently been extended towards $\nu A$ scattering~\cite{Simo:2016ikv}. 
Other approaches have also been developed. In ab-initio calculations of sum rules for neutral currents on $^{12}$C \cite{Lovato:2013cua,Lovato:2015qka}, the nuclear correlations and the MEC contributions are inherently taken into account. The authors conclude that the presence of two-body currents significantly influences the nuclear responses and sum rules, even at QE kinematics. Recent work on electron scattering by Benhar \textit{et al.}~\cite{Benhar:2015ula} and Rocco \textit{et al.}
\cite{Rocco:2015cil} have generalized the formalism based on a factorization ansatz and nuclear spectral functions to treat transition matrix elements involving two-body currents. 

In this paper, we present a model which goes beyond the IPM by implementing SRCs in the nuclear wave functions. 
This work is a first step in an extension towards the weak sector of the model developed by the Ghent group, which accounts for MEC as well as SRCs, for photoinduced \cite{Ryckebusch:1993tf} and electroinduced \cite{Ryckebusch:1997gn,Janssen:1999xy} 1p1h and 2p2h reactions. The model describes exclusive $^{16}$O$(e,e^\prime
\textnormal{pp})$ 
\cite{Starink:2000aa,Ryckebusch:2003tu}, semi-exclusive $^{16}$O$(e,e^\prime \textnormal{p})$ \cite{Fissum:2004we,Iodice:2007mn} as well as inclusive $^{12}$C$(e,e^\prime)$ and $^{40}$Ca$(e,e^\prime)$ \cite{VanderSluys:1993cv} scattering with a satisfactory accuracy.
Several groups studied two-body effects in exclusive $eA$ interactions \cite{Giusti:1997pa, Anguiano:2002sg,Anguiano:2002jz}, but so far have not presented results for weak interactions.
The continuum and bound-state wave functions in this work are computed using a Hartree-Fock (HF) method with the same Hamiltonian. This approach guarantees that the initial and final nuclear states are orthogonal. This is of great importance in view of the evaluation of multinucleon corrections to the cross section. The influence of SRCs is examined by calculating transition matrix elements of the one-body nuclear current between correlated nuclear states. Our approach translates into the calculation of transition matrix elements of an effective operator, which consists of a one- and a two-body part, between uncorrelated nuclear wave functions. The influence of the central, tensor and spin-isospin correlations are studied. 

\begin{figure}[h]
  \centering
  \includegraphics{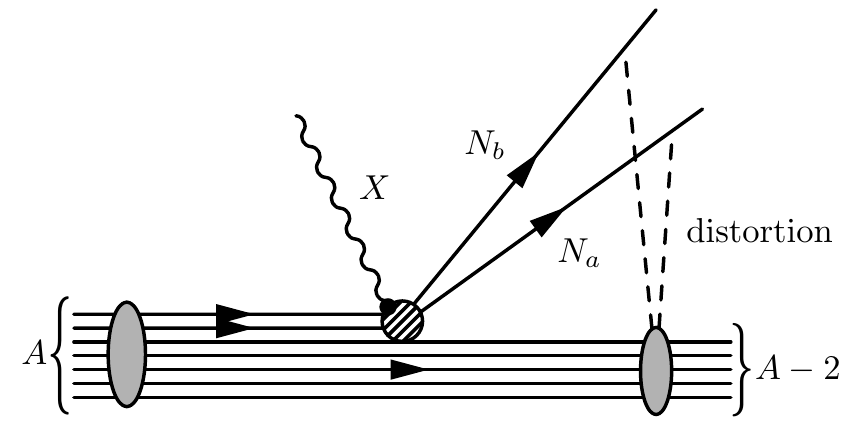}
  \caption{Graphical presentation of a 2p2h-excitation induced by SRCs (dashed area) with distortion effects (dashed lines) from the $A-2$ spectator nucleons. The boson $X$ can be either a $\gamma^*$ or a $W^+$ in this work.}
  \label{fig:model}
\end{figure}

In this work, we will refer to the double differential cross section as a function of the energy transfer and lepton scattering angle 
as the inclusive quasielastic $A(\nu_\mu,\mu^-)$ cross section. Both one-nucleon ($1N$) and two-nucleon ($2N$) knockout contribute, as do other processes, such as meson production, which are not included in this work. A second topic addressed in this paper is that of exclusive $A(\nu_\mu,\mu^- N_a N_b)$ reactions, where next to the scattered $\mu^-$, two outgoing nucleons are detected. 
Up to now the theoretical papers studying multinucleon excitations in $\nu A$ scattering \cite{Martini:2009uj,Martini:2010ex,Amaro:2010sd,Nieves:2011pp,Bodek:2011ps,Martini:2011wp,Nieves:2011yp,Amaro:2011aa,Lalakulich:2012ac,Nieves:2013fr,Martini:2013sha,Gran:2013kda,Martini:2014dqa,Megias:2014qva,Ericson:2015cva,Ivanov:2015aya,Martini:2016eec} have considered only inclusive processes.
The semi-exclusive $A(\nu_\mu,\mu^- N)$ reactions detect only one of the outgoing nucleons, but the residual nuclear system is excited above the $2N$ emission threshold. From the experimental side, the ArgoNeuT collaboration recently published the first results of exclusive neutrino interactions, where a clear back-to-back knockout signal was detected in a subset of the events \cite{Acciarri:2014gev}. Experiments using liquid argon detectors such as MicroBooNE \cite{microboone} and DUNE \cite{Adams:2013qkq} or scintillator trackers
such as MINERvA \cite{minerva} and NOvA \cite{nova} will also be able to measure exclusive cross sections.

This paper is organized as follows. In Sect.~\ref{sec:for} we describe the formalism used to account for SRCs in lepton-nucleus scattering. In Sect.~\ref{sec:one} we address $^{12}$C$(e,e^\prime)$ $1N$ knockout and describe the influence of SRCs. 
  In Sect.~\ref{sec:two} $2N$ knockout cross sections are studied. 
  First the exclusive $^{12}$C$(\nu_\mu,\mu^- N_a N_b)$ cross sections are examined, which show a clear back-to-back dominance. 
  Next, the exclusive $2N$ knockout cross section is used to calculate the semi-exclusive and the inclusive cross sections. The inclusive $^{12}$C$(e,e^\prime)$ cross section with $1N$ and $2N$ knockout is presented as a benchmark.
  Finally, in Sect.~\ref{sec:res}, we present results for inclusive $^{12}$C$(\nu_\mu,\mu^-)$ cross sections.

\section{Short-range correlations and nuclear currents}\label{sec:for}
Different techniques to correct IPM wave functions for correlations have been developed over the years. We follow the approach outlined in Refs.~\cite{Ryckebusch:1997gn,Janssen:1999xy,Vanhalst:2012ur,Vanhalst:2014cqa}. Upon calculating transition matrix elements in an IPM, the nuclear wave functions are written as Slater determinants $|\Phi\rangle$. The correlated wave functions $|\Psi\rangle$ are constructed by applying a many-body correlation operator
$\widehat{\mathcal{G}}$ to the uncorrelated wave functions $|\Phi\rangle$, 
\begin{align}
  |\Psi\rangle=\frac{1}{\sqrt{\mathcal{N}}}\widehat{\mathcal{G}}|\Phi\rangle,
\end{align}
with $\mathcal{N} =  \langle \Phi | \widehat{\mathcal{G}}^\dagger \widehat{\mathcal{G}} |\Phi \rangle$ the normalization constant. In determining $\widehat{\mathcal{G}}$, one is guided by the basic features of the one-boson exchange nucleon-nucleon force which contains many terms. Its short-range part, however, is dominated by the central $(c)$, tensor $(t\tau)$ and spin-isospin
$(\sigma \tau)$ component. To a good approximation, $\widehat{\mathcal{G}}$
can be written as
\begin{align}
  \widehat{\mathcal{G}}\approx\widehat{\mathcal{S}} \left( \prod^A_{i<j} \left[ 1 + \widehat{l}(i,j) \right] \right), 
\end{align} 
with $\widehat{\mathcal{S}}$ the symmetrization operator and
\begin{align}
  \widehat{l}(i,j) 
  =&-g_c(r_{ij}) + f_{\sigma \tau}(r_{ij}) \left( \vec{\sigma}_i \cdot \vec{\sigma}_j \right) \left( \vec{\tau}_i \cdot \vec{\tau}_j \right) \\ &+ f_{t\tau}(r_{ij})\widehat{S}_{ij} \left( \vec{\tau}_i \cdot \vec{\tau}_j \right),
\end{align}
where $r_{ij} = |\vec{r}_i - \vec{r}_j|$ and $\widehat{S}_{ij}$ is the tensor operator
\begin{align}
  \widehat{S}_{ij}=\dfrac{3}{r^2_{ij}}(\vec{\sigma}_i \cdot \vec{r}_{ij})(\vec{\sigma}_j \cdot \vec{r}_{ij})-(\vec{\sigma}_i \cdot \vec{\sigma}_j).
\end{align}
This paper uses the central correlation function $g_c(r_{ij})$ by Gearhaert and Dickhoff \cite{gearhaert} and the tensor $f_{t \tau}(r_{ij})$ and spin-isospin correlation functions $f_{\sigma \tau}(r_{ij})$ by Pieper \textit{et al.}~\cite{Pieper:1992gr}. For small internucleon distances, $f_{t\tau}$ and $f_{\sigma \tau}$ 
are considerably weaker than $g_c$. At medium inter-nucleon distances ($r_{ij} \gtrsim 3 \textnormal{ fm}$), $\widehat{l}(r_{ij}) \rightarrow 0$. In momentum space $f_{t\tau}$ dominates for relative momenta $200 - 400$ MeV/c \cite{Janssen:1999xy}.\\
Transition matrix elements between correlated states $|\Psi\rangle$ can be written as matrix elements between uncorrelated states $|\Phi\rangle$, whereby the effect of the SRCs is implemented as an effective transition operator \cite{Ryckebusch:1997gn,Janssen:1999xy,Vanhalst:2014cqa}
\begin{align}
  \langle \Psi_\textnormal{f} | \widehat{J}_\mu^\textnormal{nucl} | \Psi_\textnormal{i} \rangle = \frac{1}{\sqrt{\mathcal{N}_\textnormal{i}\mathcal{N}_\textnormal{f}}} \langle \Phi_\textnormal{f} | \widehat{J}_\mu^{\textnormal{eff}} | \Phi_\textnormal{i} \rangle.
\end{align}
In the IA, the many-body nuclear current can be written as a sum of one-body operators
\begin{align}
  \widehat{J}^\textnormal{nucl}_\mu = \sum_{i=1}^A \widehat{J}^{[1]}_\mu(i). 
  \label{}
\end{align}
The effective nuclear current, which accounts for SRCs, can be written as
\begin{align} 
  \widehat{J}_\mu^{\textnormal{eff}} &\approx \sum_{i=1}^A \widehat{J}_\mu^{[1]}(i) \nonumber \\ &+ \sum_{i<j}^A \widehat{J}_\mu^{[1],\textnormal{in}}(i,j) + \left[ \sum_{i<j}^A \widehat{J}_\mu^{[1],\textnormal{in}}(i,j) \right]^\dagger,
\label{eq:jeff}
\end{align}
with
\begin{align}
  \widehat{J}^{[1],\textnormal{in}}_\mu(i,j) &= \left[ \widehat{J}^{[1]}_\mu(i) + \widehat{J}^{[1]}_\mu(j) \right] \widehat{l}(i,j).\label{eq:effop}
\end{align}
The effective operator consists of one- and two-body terms. The superscript 'in' refers to initial-state correlations. In the expansion of the effective operator, only terms that are linear in the correlation operators are retained. In Ref.~\cite{Vanhalst:2014cqa} it is argued that this approximation accounts for the majority of the SRC effects.

\section{SRC corrections to inclusive one-nucleon knockout}\label{sec:one}
\begin{figure}[ht]
  \centering
  \includegraphics[scale=0.85]{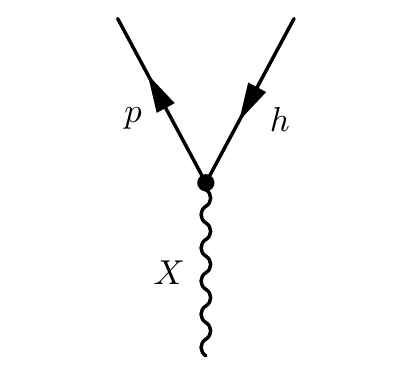}
  \centering
  \includegraphics[scale=0.85]{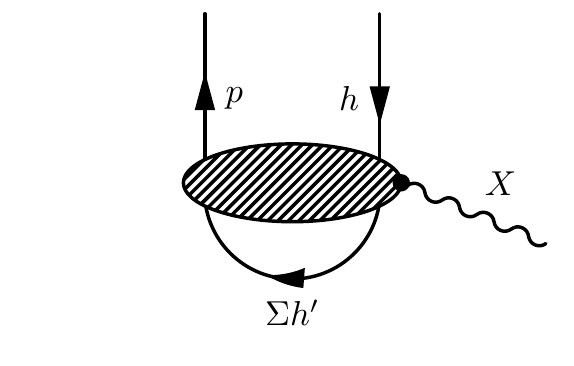}
  \caption{Diagrams considered in the 1p1h calculations reported in this paper. The left diagram shows the 1p1h channel in the IA and the right diagram shows the SRC corrections (dashed oval).}\label{fig:one}
\end{figure}
\begin{figure*}
  \centering
  \includegraphics[angle=0,width=0.95\textwidth]{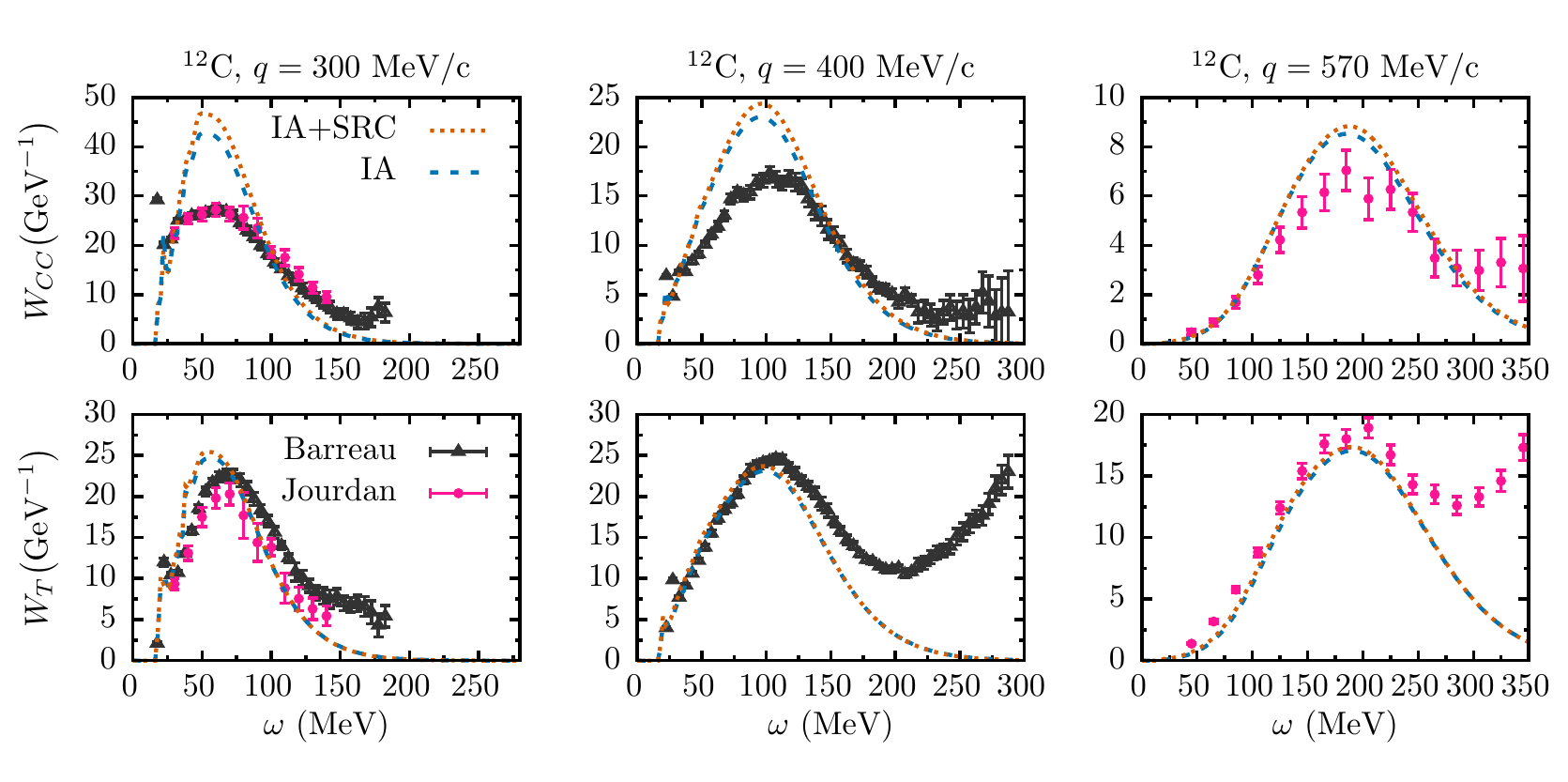}
  \caption{(Color online) The $\omega$ dependence of the longitudinal ($W_{CC}$) and transverse ($W_T$) responses for the 1p1h contribution to $^{12}$C$(e,e')$. Results are shown for three values of the momentum transfer $q$. The data is from Refs.~\cite{Barreau:1983ht,Jourdan:1996np}.
}
\label{fig:iasrcelecq}
\end{figure*}
In this section we describe electron and charged-current (CC) muon-neutrino ($\nu_\mu$) induced $1N$ knockout 
  \begin{align}
    e(E_e,\vec{k}_e) + A &\rightarrow e'(E_{e'},\vec{k}_{e'}) + (A-1)^* + N(E_N,\vec{p}_N) \nonumber \\
    \nu_\mu(E_{\nu_\mu},\vec{k}_{\nu_\mu}) + A &\rightarrow \mu(E_\mu,\vec{k}_\mu) + (A-1)^* + N(E_N,\vec{p}_N). \nonumber
  \end{align}
  Throughout this work we will refer to the initial lepton as $l$ and the final state lepton as $l'$. The four-momentum transfer, $q^\mu=(\omega,\vec{q})$, is
  \begin{align}
    \omega &= E_l - E_{l'}, & \vec{q} &= \vec{k}_l - \vec{k}_{l'},
  \end{align}
  and $Q^2 = \vec{q}^{\,2} - \omega^2$. In the $1N$ knockout channel, we calculate the inclusive responses and integrate over $\Omega_N$. The double differential $A(e,e^\prime)$ cross section is written as
  \begin{align}
    \dfrac{\mathrm{d}\sigma}{\mathrm{d}E_{e'} \mathrm{d}\Omega_{e'}} = \sigma^\mathrm{Mott} \bigl[& v^e_{L} W_{CC} + v^e_T W_T \bigr].\label{eq:elecone}
  \end{align}
  For CC $A(\nu_\mu,\mu^-)$ interactions, one has
  \begin{align}
    \dfrac{\mathrm{d}\sigma}{\mathrm{d}E_\mu \mathrm{d}\Omega_\mu} = \sigma^{W} \zeta \bigl[& v_{CC} W_{CC} + v_{CL} W_{CL} + v_{LL} W_{LL} \nonumber \\
    &+ v_T W_T  \mp v_{T'} W_{T'} \bigr],\label{eq:neutone}
  \end{align}
  the $-(+)$ sign refers to neutrino(antineutrino) scattering. The prefactors are defined as
  \begin{align}
    \sigma^\mathrm{Mott} &= \left( \frac{\alpha \cos(\theta_{e'}/2)}{2E_e\sin^2(\theta_{e'}/2)}\right)^2, \\ 
    \sigma^W &= \left( \frac{G_F \cos(\theta_c) E_\mu}{2\pi}\right)^2,
  \end{align}
  with $\alpha$ the fine-structure constant, $\theta_{e'}$ the electron scattering angle, $G_F$ the Fermi constant, $\theta_c$ the Cabibbo angle and the kinematic factor $\zeta$
  \begin{align}
    \zeta=\sqrt{1-\frac{m_\mu^2}{E_\mu^2}}.
  \end{align}
  The functions $v$ contain the lepton kinematics and the response functions $W$ the nuclear dynamics. The $W$ are defined as products of transition matrix elements 
  $\mathcal{J}_\lambda$
  \begin{align}
    \mathcal{J}_\lambda = \langle \Psi_\textnormal{f} | \widehat{J}_\lambda(q) | \Psi_\textnormal{i} \rangle. \label{eq:currents}
  \end{align}
  Here, $| \Psi_\textnormal{f}\rangle$ and $|\Psi_\textnormal{i}\rangle$ refer to the final and initial correlated nuclear state and $\widehat{J}_\lambda$ are the spherical components of the nuclear four-current in the IA. The results presented in this work consider $^{12}$C as target nucleus. For $^{12}$C$(e,e^\prime)$  two 1p1h final states are accessible 
  \begin{align}
    | \Psi_\textnormal{f} \rangle^{1p1h} = | ^{11}\textnormal{C},\textnormal{n} \rangle ,\, | ^{11}\textnormal{B},\textnormal{p} \rangle,
  \end{align}
  while for CC neutrino scattering only one 1p1h final state is accessible
  \begin{align}
    | \Psi_\textnormal{f} \rangle^{1p1h} = | ^{11}\textnormal{C},\textnormal{p} \rangle .
  \end{align}
  The expressions for the kinematic factors and the response functions are given in Appendix \ref{app:cro}. 
  As explained in Sect.~\ref{sec:for}, we replace the one-body nuclear current $\widehat{J}_\lambda$ in (\ref{eq:currents}) with the effective nuclear current $\widehat{J}_\lambda^\textnormal{eff}$, which accounts for SRCs. This results in a coherent sum of a one- and a two-body contribution to the $\mathcal{J}_\lambda$
  \begin{align}
    \mathcal{J}_\lambda \approx \mathcal{J}_\lambda^{(1)} + \mathcal{J}_\lambda^{(2)},
  \end{align}
  where
  \begin{align}
    \mathcal{J}_\lambda^{(1)} &= \sum_{i=1}^A \langle \Phi_\textnormal{f}^{(A-1)}(J_R,M_R);\vec{p}_N m_s |\widehat{J}_\lambda^{[1]}(i) | \Phi_\textnormal{gs} \rangle,\label{eq:onecur} \\
    \mathcal{J}_\lambda^{(2)} &= \sum_{i<j}^A \langle \Phi_\textnormal{f}^{(A-1)}(J_R,M_R);\vec{p}_N m_s |\widehat{J}_\lambda^{[1],\textnormal{in}}(i,j) | \Phi_\textnormal{gs} \rangle \nonumber \\
    &+ \sum_{i<j}^A \langle \Phi_\textnormal{f}^{(A-1)}(J_R,M_R);\vec{p}_N m_s | \left[ \widehat{J}_\lambda^{[1],\textnormal{in}}(i,j) \right]^\dagger | \Phi_\textnormal{gs} \rangle \label{eq:twocur},
  \end{align}
  with $| \Phi_\textnormal{gs} \rangle$ the ground-state Slater determinant of the target nucleus. The bra states have an on-shell nucleon with momentum $\vec{p}_N$ and spin $m_s$ and a residual $A-1$ nucleus with quantum numbers $J_R,M_R$, which can either be the ground state or a low lying excited state.

  We work in the so-called spectator approach (SA), where the nucleon absorbing the boson is the one that becomes asymptotically free. The nucleon in the continuum, however, is still under influence of the potential of the $A-1$ system, the outgoing waves are no plane waves.   
  This distortion effect of the residual nuclear system on the continuum nucleon is accounted for by computing the continuum and bound-state wave functions using the same potential \cite{Ryckebusch:1994zz}. 
  The wave functions are constructed through a HF calculation with an effective Skyrme-type interaction \cite{Waroquier:1986mj}. The single-particle wave functions are calculated in a nonrelativistic framework. Relativistic corrections are implemented in an effective fashion as explained in Refs.~\cite{Amaro:1995kk,Amaro:2005dn}. 
  This can be achieved by the following substitution for $\omega$ in the computation of the outgoing nucleon wave functions
  \begin{align}
     \omega \rightarrow \omega \left( 1 + \frac{\omega}{2 m_N} \right) ,
  \end{align}
  with $m_N$ the nucleon mass. 
The HF wave functions used in this model successfully describe the low energy side of the quasielastic $\nu A$ and $\overline{\nu} A$ cross sections using a continuum random phase approximation (CRPA) with relativistic lepton kinematics \cite{Jachowicz:1998fn,Jachowicz:2002rr,Pandey:2013cca,Pandey:2014tza}.

  When adopting a multipole expansion, the calculation of the amplitudes (\ref{eq:onecur}) can be reduced to the computation of 1p1h matrix elements of the form
  \begin{align} 
    \langle ph^{-1} | \widehat{\mathcal{O}}_{JM}^{(1)}(q) | \Phi_0 \rangle =& (-1)^{j_p-m_p} 
    \begin{pmatrix} j_p & J & j_h \\ -m_p & M & m_h \end{pmatrix} \nonumber \\ & \times \langle p || \widehat{\mathcal{O}}_J^{(1)}(q) || h \rangle, \label{eq:1p1h}
  \end{align}
  with $| \Phi_0 \rangle$ the single-particle vacuum and $\widehat{\mathcal{O}}_{JM}^{(1)}$ a multipole operator as defined in Appendix~\ref{app:mat}. The evaluation of the two-body part of the matrix elements (\ref{eq:twocur}) reduces to ($\widehat{J} \equiv \sqrt{ 2 J + 1}$)
  \begin{align}
    \langle ph^{-1} | \widehat{\mathcal{O}}_{JM}^{(2)}(q) | \Phi_0 \rangle =& \sum_{h'} \sum_{J_1 J_2} \widehat{J}_1 \widehat{J}_2  (-1)^{-j_p+j_h'-J_2-M} \nonumber \\ 
    \times& \begin{pmatrix} j_p & J & j_h \\ m_p & -M & -m_h \end{pmatrix} 
    \begin{Bmatrix} j_p & J & j_h \\ J_2 & j_h' & J_1 \end{Bmatrix} \nonumber \\ \times & \langle ph';J_1 || \widehat{\mathcal{O}}_J^{(2)}(q) || hh';J_2 \rangle_\textnormal{as} \label{eq:red1p1h},
  \end{align}
  with $\widehat{\mathcal{O}}_{JM}^{(2)}$ a two-body operator, defined as in Eq.~(\ref{eq:effop}). The sum $\sum_{h'}$ extends over all occupied single-particle states of the target nucleus. The antisymmetrized reduced matrix element is defined as
  \begin{align}
    \langle ab ; J_1 || \widehat{\mathcal{O}}_{J}^{(2)}(q) || cd ; J_2 \rangle_\textnormal{as} = &\langle ab ; J_1 || \widehat{\mathcal{O}}_{J}^{(2)}(q) || cd ; J_2 \rangle \nonumber \\ 
    - (-1)^{j_c + j_d-J_2}& \langle ab ; J_1 || \widehat{\mathcal{O}}_{J}^{(2)}(q) || dc ; J_2 \rangle. 
  \end{align}
  The reduced matrix elements accounting for correlations are discussed in Appendix~\ref{app:mat}. The diagrams corresponding with the matrix elements in Eqs.~(\ref{eq:1p1h}) and (\ref{eq:red1p1h}) are shown in Fig.~\ref{fig:one}.  

  The influence of SRC currents on the 1p1h $^{12}$C$(e,e^\prime)$ responses is shown in Fig.~\ref{fig:iasrcelecq} and compared with data. The form factors used in the electron scattering calculations are the standard dipole form factors and a Galster parameterization for the neutron electric form factor \cite{Galster:1971kv}. The predictions are compared with Rosenbluth separated cross section data for a fixed momentum transfer. The IA calculations overestimate the longitudinal responses, while the transverse responses are
  slightly underestimated for $\omega$-values beyond the QE-peak. The differences can be attributed to long-range correlations \cite{Pandey:2014tza}. These results are
  in-line with other predictions using similar approaches \cite{Jourdan:1996np,Amaro:1999nj}. The two-body corrections from SRCs in the 1p1h channel result in
a small increase of the longitudinal and a marginal increase of the transverse response function.

  \section{SRC contribution to two-nucleon knockout}\label{sec:two}
  For $2N$ knockout, the following interactions are considered
  \begin{align}
    e(E_e,\vec{k}_e) + A &\rightarrow e'(E_{e'},\vec{k}_{e'}) + (A-2)^* \nonumber \\
    &\quad + N_a(E_a,\vec{p}_a) + N_b(E_b,\vec{p}_b),\\
    \nu_\mu(E_{\nu_\mu},\vec{k}_{\nu_\mu}) + A &\rightarrow \mu(E_{\mu},\vec{k}_{\mu}) + (A-2)^* \nonumber \\
    &\quad + N_a(E_a,\vec{p}_a) + N_b(E_b,\vec{p}_b).
  \end{align}
Electroinduced $2N$ knockout has three possible final states,
  \begin{align}
    | \Psi_\textnormal{f} \rangle^{2p2h} = | ^{10}\textnormal{Be},\textnormal{pp} \rangle ,\, | ^{10}\textnormal{B},\textnormal{pn} \rangle,\, | ^{10}\textnormal{C},\textnormal{nn} \rangle,
  \end{align}
  while CC neutrino reactions have two possible final states
  \begin{align}
    | \Psi_\textnormal{f} \rangle^{2p2h} = | ^{10}\textnormal{B},\textnormal{pp} \rangle ,\, | ^{10}\textnormal{C},\textnormal{pn} \rangle.
  \end{align}
  The two-body transition matrix elements are given by
  \begin{align}
    \mathcal{J}_\lambda = \sum_{i<j}^A\langle \Phi_\textnormal{f}^{(A-2)}(J_R,M_R);\vec{p}_a m_{a} ; \vec{p}_b m_{b} | \widehat{J}^{[1],\textnormal{in}}_\lambda(i,j) | \Phi_\textnormal{gs} \rangle & \nonumber \\
    + \sum_{i<j}^A\langle \Phi_\textnormal{f}^{(A-2)}(J_R,M_R);\vec{p}_a m_{a} ; \vec{p}_b m_{b} | \left[ \widehat{J}^{[1],\textnormal{in}}_\lambda(i,j) \right]^\dagger | \Phi_\textnormal{gs} \rangle &,\label{eq:twotwo}
  \end{align}
  where two outgoing nucleons are created along with the residual $A-2$ nucleus. Only the two-body part of the effective nuclear current contributes to the $2N$ knockout cross section. In $2N$ knockout from finite nuclei, we follow the same approach as in the $1N$ knockout calculations, adopt the SA and neglect the mutual interaction between
  the outgoing particles.

\begin{figure}
    \centering
    \includegraphics[scale=0.85,trim= 1cm 1cm 1cm 0cm, clip]{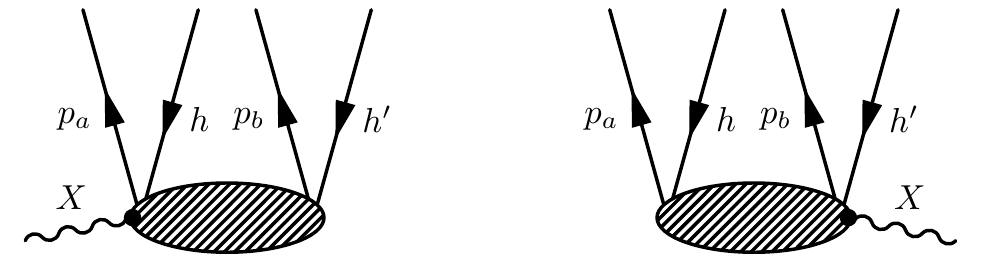}
    \caption{Diagrams considered in the $2N$ knockout calculations.}
    \label{fig:two}
  \end{figure}

  The diagrams considered in the $2N$ knockout calculations presented in this paper are shown in Fig.~\ref{fig:two}. In the adopted multipole expansion, the calculation of the transition amplitudes (\ref{eq:twotwo}) is reduced to the calculation of 2p2h matrix elements of the form
  \begin{align}
    \langle p_a p_b& (hh')^{-1} | \widehat{O}_{JM}^{(2)}| \Phi_0 \rangle = \sum_{J_1 M_1} \sum_{J_R M_R} \frac{(-1)^{J_R + M_R + 1}}{\widehat{J}_1} \nonumber \\
    &\times \langle j_a m_{j_a} , j_b m_{j_b} | J_1 M_1 \rangle \langle J_R -M_R, J M | J_1 M_1 \rangle \nonumber \\
    &\times \langle j_h m_h, j_h' m_h' | J_R M_R \rangle \nonumber \\
    &\times  \langle p_a p_b ; J_1 || \widehat{O}_{J}^{(2)} || hh';J_R \rangle_\textnormal{as}.\label{eq:red2p2h}
  \end{align}
    Note that the reduced matrix elements in Eqs.~(\ref{eq:red1p1h}) and (\ref{eq:red2p2h}) have exactly the same structure. All the differential cross sections for $2N$ knockout presented below, are obtained by incoherently adding the possible final states. 
    With $^{12}$C as a target nucleus, $2N$ knockout from all possible shell combinations is considered.
  \subsection{Exclusive $2N$ knockout cross section}
  The exclusive $A(e,e' N_a N_b)$ cross section in the lab frame, can be written as a function of four response functions
  \begin{align}
    \dfrac{\mathrm{d}\sigma}{\mathrm{d}E_{e'} \mathrm{d}\Omega_{e'} \mathrm{d}T_a \mathrm{d}\Omega_a\mathrm{d}\Omega_b} = \sigma^\mathrm{Mott} f_{rec}^{-1} \nonumber \\
    \times \bigl[ v^e_{L} W_{CC} + v^e_T W_T + v^e_{TT} W_{TT} + v^e_{TL} W_{TC}  \bigr],\label{eq:electwo}
  \end{align}
  with recoil factor
  \begin{align}
    f_{rec} = \left| 1 + \frac{E_b}{E_{A-2}}\left(1 - \frac{ \vec{p}_b \cdot \left( \vec{q} - \vec{p}_a \right) }{p_b^2} \right) \right|.
  \end{align}
  Ten response functions contribute to $A(\nu_\mu,\mu^- N_a N_b)$ reactions,
  \begin{align}
    \dfrac{\mathrm{d}\sigma}{\mathrm{d}E_\mu \mathrm{d}\Omega_\mu \mathrm{d}T_a \mathrm{d}\Omega_a\mathrm{d}\Omega_b} &	= \sigma^{W} \zeta f_{rec}^{-1} \nonumber \\
    \times \bigl[ v_{CC} W_{CC} + v_{CL} W_{CL} &+ v_{LL} W_{LL} + v_T W_T \nonumber \\ 
      + v_{TT} W_{TT} &+ v_{TC} W_{TC} + v_{TL} W_{TL} \nonumber \\
    \mp (v_{T'} W_{T'} + v_{TC'} W_{TC'} &+ v_{TL'} W_{TL'}) \bigr].\label{eq:neuttwo}
  \end{align}
  The kinematic functions $v$ and response functions $W$ are defined in Appendix~\ref{app:cro} and $T_a$ refers to the kinetic energy of particle $a$. The azimuthal information of the emitted nucleons is contained in $W_{TT}, W_{TC}, W_{TL}, W_{TC'}$ and $W_{TL'}$, while all the response functions depend on $\theta_a$ and $\theta_b$.

\begin{figure}
    \centering
    \includegraphics[width=0.95\columnwidth]{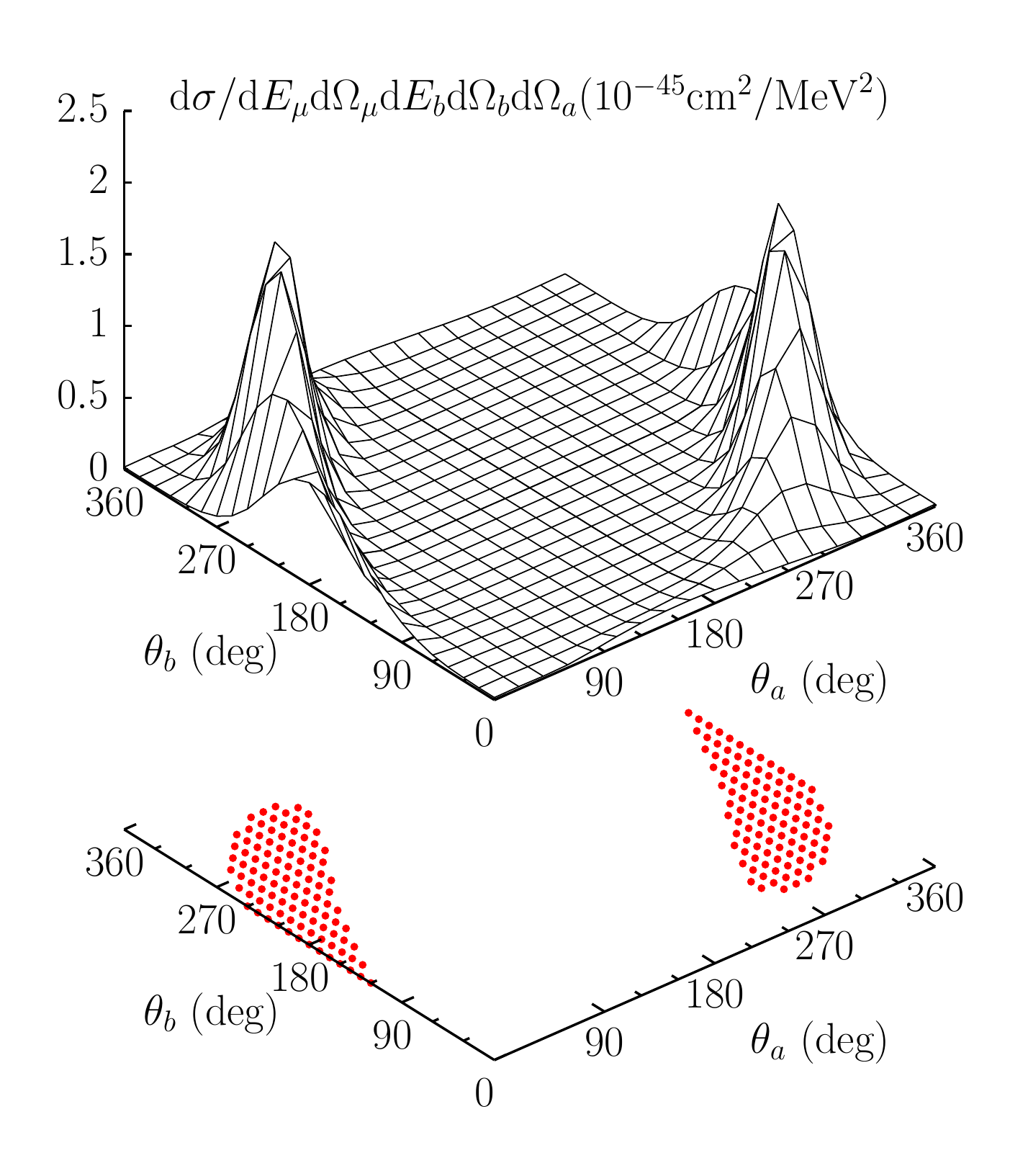}
    \caption{(Color online) The $^{12}$C$(\nu_\mu,\mu^- N_a N_b)$ cross section ($N_a =$ p, $N_b = $ p$^\prime$, n) at $E_{\nu_\mu} = 750$ MeV, $E_\mu=550$ MeV, $\theta_\mu=15^\circ$ and $T_\textrm{p}=50$ MeV for in-plane kinematics. The bottom plot shows the ($\theta_a,\theta_b)$ regions with $P_{12} < 300$ MeV/c.
    \label{fig:srcexcl}
  }
\end{figure}

  In Fig.~\ref{fig:srcexcl} the result of an exclusive $^{12}$C$(\nu_\mu,\mu^- N_a N_b)$ cross section is shown ($N_a =$ p, $N_b = $ p$^\prime$, n). We consider in-plane kinematics, with both nucleons emitted in the lepton scattering plane. A striking feature of the cross section is the dominance of back-to-back nucleon knockout, reminiscent of the 'hammer events' seen by the ArgoNeuT
  collaboration \cite{Acciarri:2014gev}. This feature is independent of the interacting
  lepton and the type of two-body currents, whether they be SRCs or MECs (see Refs.~\cite{Ryckebusch:1993tf,Ryckebusch:1997gn,Janssen:1999xy}). 

  For $2N$ knockout reactions, momentum conservation can be written as
  \begin{align}
    \vec{P}_{12} + \vec{q} = \vec{p}_a + \vec{p}_b,
    \label{eq:miss}
  \end{align}
  where $\vec{P}_{12}$ is the initial center-of-mass (c.o.m.) momentum of the pair. Referring to Fig.~\ref{fig:srcexcl}, it is clear that most strength is residing in a region with $P_{12} < 300$ MeV/c. This behavior can be understood in a factorized model \cite{Ryckebusch:1996wc,Ryckebusch:1997gn,Colle:2013nna,Colle:2015ena}, that shows that the SRC dominated part of the $2N$ knockout cross section is proportional to the c.o.m.~distribution of close-proximity pairs.
\subsection{Semi-exclusive $2N$ knockout cross section}
In this section, we compute the contribution of exclusive $2N$ knockout $A(l,l^\prime N_a N_b)$ to the semi-exclusive $A(l,l^\prime N_a)$ cross section with the residual nuclear system $(A-1)^*$ excited above the $2N$ emission threshold. This involves an integration over the phase space of the undetected ejected nucleons. In the case where the detected particle is a proton ($N_a =$ p, $N_b =$ p$^\prime$ or n) one has
\begin{align}
  &\dfrac{\mathrm{d}\sigma}{\mathrm{d}E_{l'} \mathrm{d}\Omega_{l'} \mathrm{d}T_\mathrm{p} \mathrm{d}\Omega_\textnormal{p}}(l,l' \mathrm{p}) \nonumber \\
  & \qquad = \int \mathrm{d}\Omega_\mathrm{p^\prime} \dfrac{\mathrm{d}\sigma}{\mathrm{d}E_{l'} \mathrm{d}\Omega_{l'} \mathrm{d}T_\mathrm{p} \mathrm{d}\Omega_\textnormal{p} \mathrm{d}\Omega_\mathrm{p^\prime}}(l,l' \mathrm{pp^\prime}) \nonumber \\ 
  & \qquad + \int \mathrm{d}\Omega_\textnormal{n} \dfrac{\mathrm{d}\sigma}{\mathrm{d}E_{l'} \mathrm{d}\Omega_{l'} \mathrm{d}T_\mathrm{p} \mathrm{d}\Omega_\textnormal{p} \mathrm{d}\Omega_\textnormal{n}}(l,l' \mathrm{pn})\label{eq:semi}.
\end{align}
One could calculate the exclusive cross section over the full phase space of the undetected nucleons and perform a numerical integration. 
We use the method outlined in \cite{Ryckebusch:1997gn} and exploit the fact that the exclusive $2N$ knockout strength resides in a well-defined part of phase space. For each particular semi-exclusive kinematic setting $(\mathrm{d} T_\textnormal{p} \mathrm{d} \Omega_\textnormal{p})$ the exclusive $(l,l'\mathrm{p} N_b)$ cross section is restricted to a small part of the phase space of the undetected particle $(\mathrm{d}\Omega_b)$, as shown in Fig.~\ref{fig:srcexcl}. In this limited part of the phase space, the momentum of the undetected particle
$\vec{p}_b$ varies very
little, which allows one to set $\vec{p}_b \approx \vec{p}_b^{\,ave}$. The average momentum ($\vec{p}_b^{\,ave}$) is determined by imposing quasi-deuteron kinematics
\begin{align}
  \vec{p}_b^{\phantom{i}ave} = \vec{q} - \vec{p}_\mathrm{p}.
\end{align}
As seen from Eq.~(\ref{eq:miss}), this average momentum is equivalent to the case where the c.o.m.~momentum of the initial pair is zero, or equivalently, where the residual nucleus has zero recoil momentum ($f_{rec}=1$). After the introduction of the average momentum, the integration over $\textnormal{d}\Omega_{\textnormal{p}^\prime}$ and $\textnormal{d}\Omega_{\textnormal{n}}$ in Eq.~(\ref{eq:semi}) can be performed analytically \cite{Ryckebusch:1997gn}.

\begin{figure}
  \centering
  \includegraphics[width=\columnwidth,trim=0cm 1.5cm 0cm 1.5cm,clip=true]{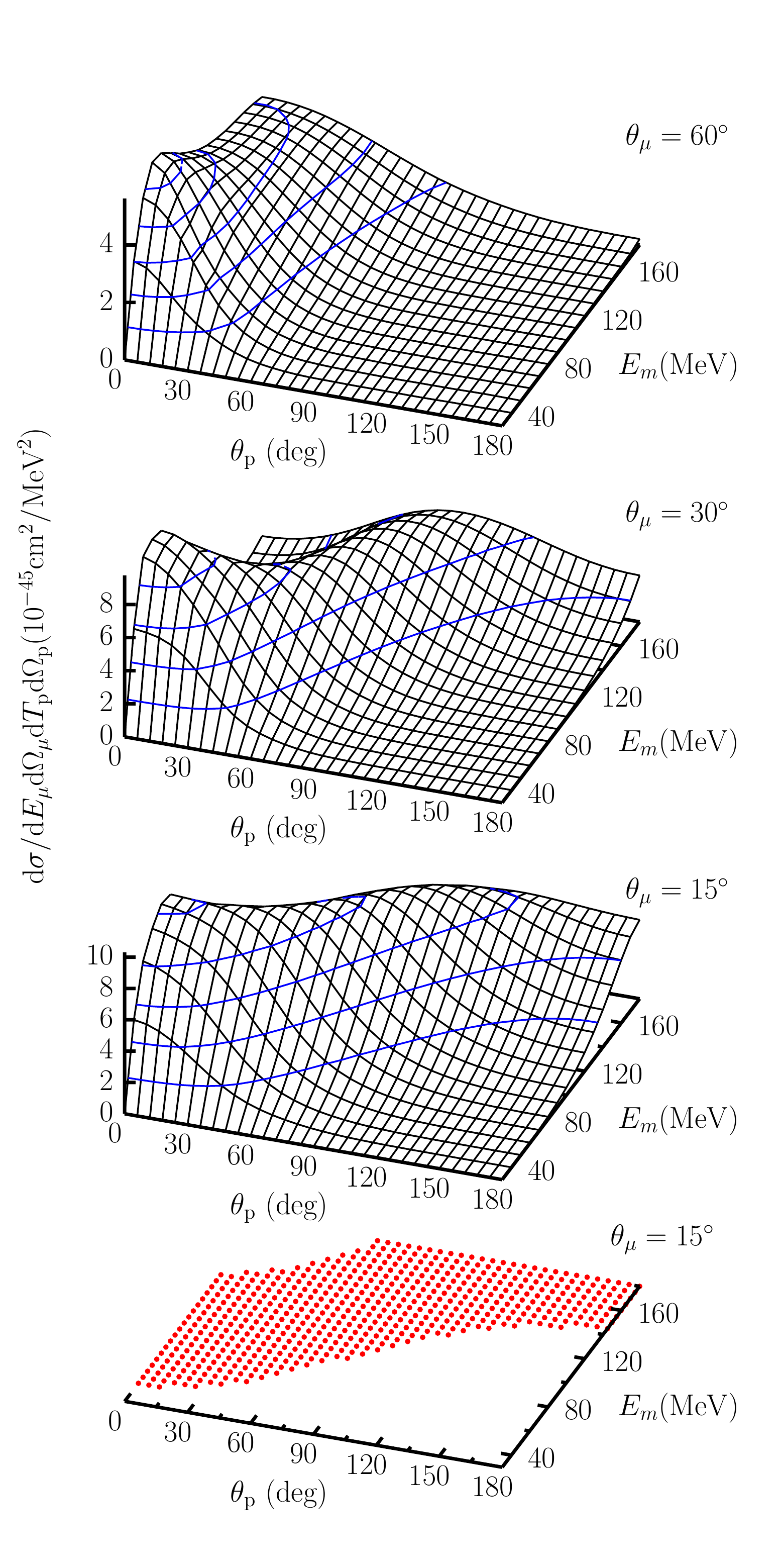}
  \caption{(Color online) Semi-exclusive $^{12}$C$(\nu_\mu,\mu^-$p$)$ cross section for $E_{\nu_\mu} = 750$ MeV, $E_\mu=550$ MeV and three muon scattering angles for in-plane kinematics ($\phi_\textnormal{p}=0^\circ)$. The bottom panel shows the ($\theta_\textnormal{p},E_m$) area with $P_{12} < 300$ MeV/c for $\theta_\mu=15^\circ$.
  \label{fig:srcsemi}
}
   \end{figure}

The results are shown in Fig.~\ref{fig:srcsemi} for three kinematics relevant for ongoing experiments. The differential cross section was studied versus missing energy $E_m = \omega-T_\textnormal{p} - T_{A-1}$ and proton angle $\theta_\textnormal{p}$ for $\phi_\textnormal{p} = 0^\circ$.

   We observe that the peak of the differential cross section shifts towards higher $E_m$ as one moves towards higher $\theta_\textnormal{p}$, where higher missing momenta are probed.   For semi-exclusive calculations, $\vec{P}_{12}$ cannot longer be reconstructed, since the angular information of one of the particles is missing. However, a Monte Carlo (MC) simulation allows one to locate the region where $P_{12}<300$ MeV/c is accessible. The bottom panel of 
   Fig.~\ref{fig:srcsemi} shows the result of such a calculation for $\theta_\mu=15^\circ$. This demonstrates that semi-exclusive cross sections are dominated by pairs with small initial c.o.m.~momentum. 
   
   Studying the different contributions separately, it can be seen that the tensor contribution is localized at small $\theta_\textnormal{p}$, whereas the contribution from the central correlations spans a wider region of the proton scattering angle, as shown for semi-exclusive $A(e.e^\prime \textnormal{p})$ in \cite{Janssen:1999xy}. This feature does not change when looking at neutrino scattering as it is a result of the fact that the central correlation function dominates at high
  ($> 400$ MeV/c) missing momenta, which are reached at larger $\theta_\textnormal{p}$. From this behavior it is expected that central correlations dominate at high $p_m$ while the tensor correlations dominate for intermediate $p_m$.
 
  It is worth remarking that at the selected kinematics, the contribution from MECs is expected to overshoot the strength attributed to SRCs \cite{Janssen:1999xy}.
   \begin{figure}
     \centering
     \includegraphics[width=0.85\columnwidth]{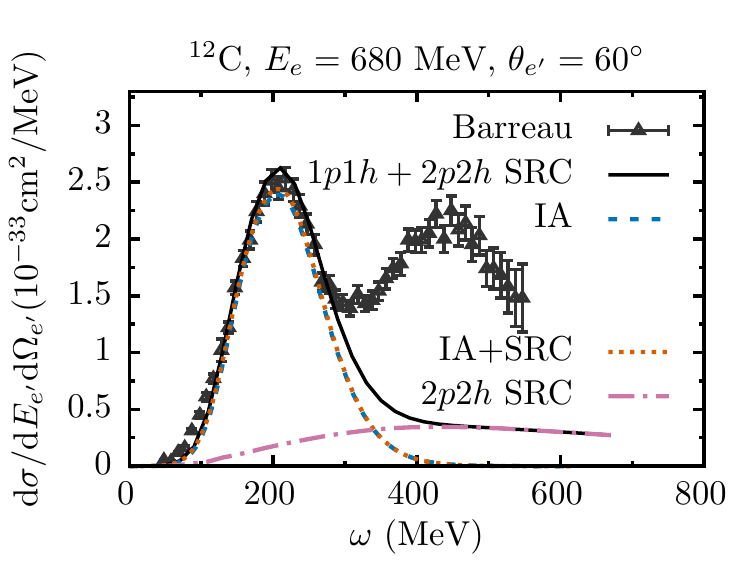}
     \caption{(Color online) The $\omega$ dependence of the $^{12}$C$(e,e^\prime)$ cross section at $E_e = 680$ MeV and $\theta_{e'}=60^\circ$. The results are compared with data from Ref.~\cite{Barreau:1983ht}.\label{fig:srcincl}
   }
 \end{figure}
 \subsection{Inclusive cross section}
 The $2N$ knockout contribution to the inclusive cross section can be calculated using the same approach. An integration over the phase space $\textnormal{d}T_\textnormal{p} \textnormal{d}\Omega_\textnormal{p}$ of the second particle is performed. For Eq.~(\ref{eq:semi}) this results in
 \begin{align}
   \frac{\mathrm{d}\sigma}{\mathrm{d}E_{l'}\mathrm{d}\Omega_{l'}} (l,l') = \int \mathrm{d}T_\textnormal{p} \mathrm{d}\Omega_\textnormal{p} \frac{\mathrm{d}\sigma}{\mathrm{d}E_{l'}\mathrm{d}\Omega_{l'} \mathrm{d}T_\textnormal{p}\mathrm{d}\Omega_\textnormal{p}} (l,l'\textnormal{p}).\label{eq:inclcross}
 \end{align}
 Performing the angular integration, it follows that five responses $\{TT,TC,TL,TC',TL'\}$ cancel since they are odd functions of $\Omega_\textnormal{p}$, the other five responses are integrated analytically. Integration over the outgoing nucleon kinetic energy $T_\textnormal{p}$ is performed numerically.

 The results of such a calculation for $^{12}$C$(e,e^\prime)$ are shown in Fig.~\ref{fig:srcincl} and compared with data. The effect of the SRCs on the 1p1h channel is very small. This is because at the selected scattering angle, the cross section is dominated by the transverse response. As discussed above, the influence of the SRCs on the transverse response was considerably smaller than in the longitudinal response in the 1p1h channel. 
 
 The 2p2h contribution to the cross sections appears as a broad background that extends into the dip region of the cross section. The majority of the strength in the 2p2h signal stems from the tensor correlations at small $\omega$, the central correlations gain in importance with growing energy transfers.

 \section{Double differential neutrino cross sections}\label{sec:res}
 \begin{figure*}[t]
   \centering
   \includegraphics[width=0.95\textwidth]{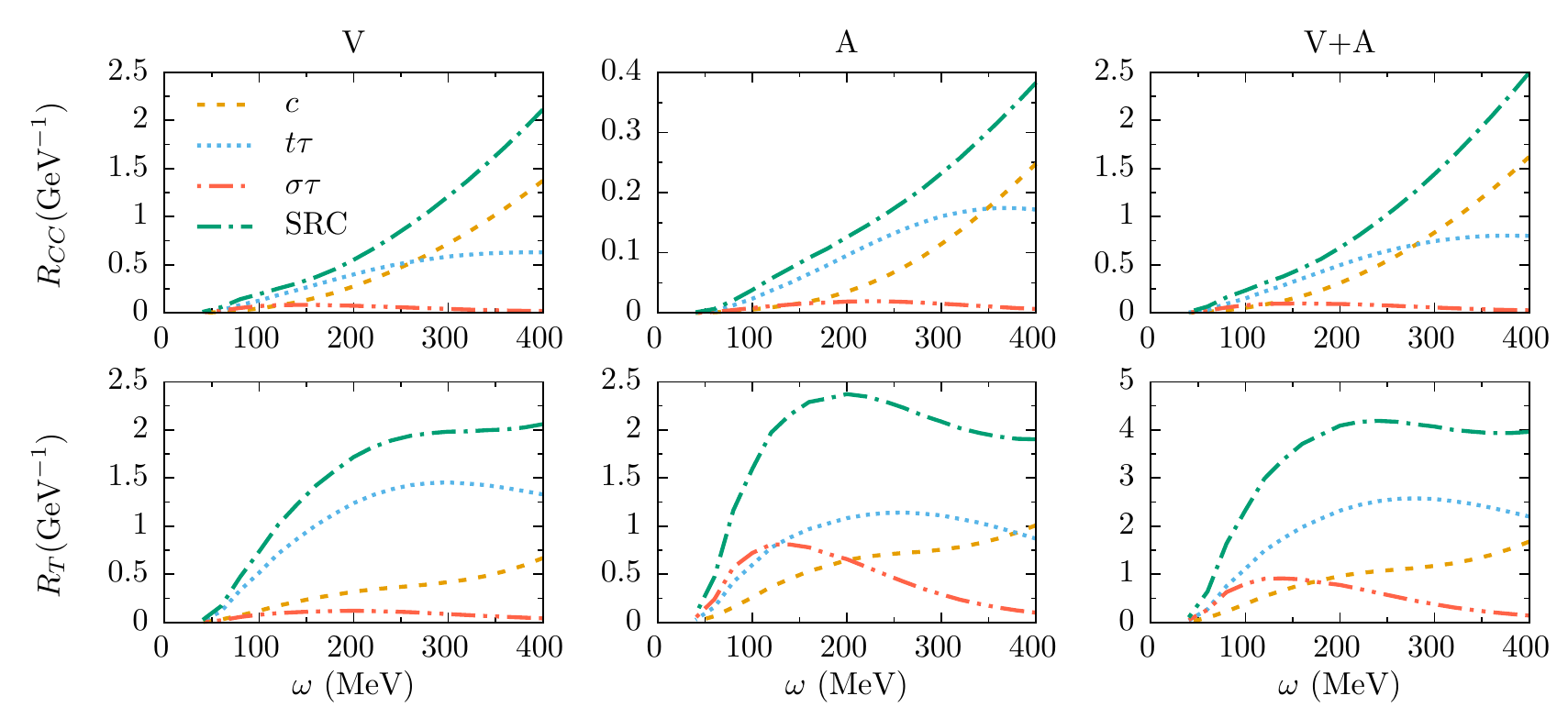}
   \caption{(Color online) The 2p2h SRC response functions $R_{CC}$ and $R_T$ for $^{12}$C$(\nu_\mu,\mu^-)$ at $q=400$ MeV/c. The contributions of the three different SRC types (SRC = $c+t\tau+\sigma\tau$) are shown for the vector (V) and axial (A) parts of the nuclear current.}
   \label{fig:neutincl4}
 \end{figure*}
 \begin{figure*}[t]
   \centering
   \includegraphics[width=0.95\textwidth]{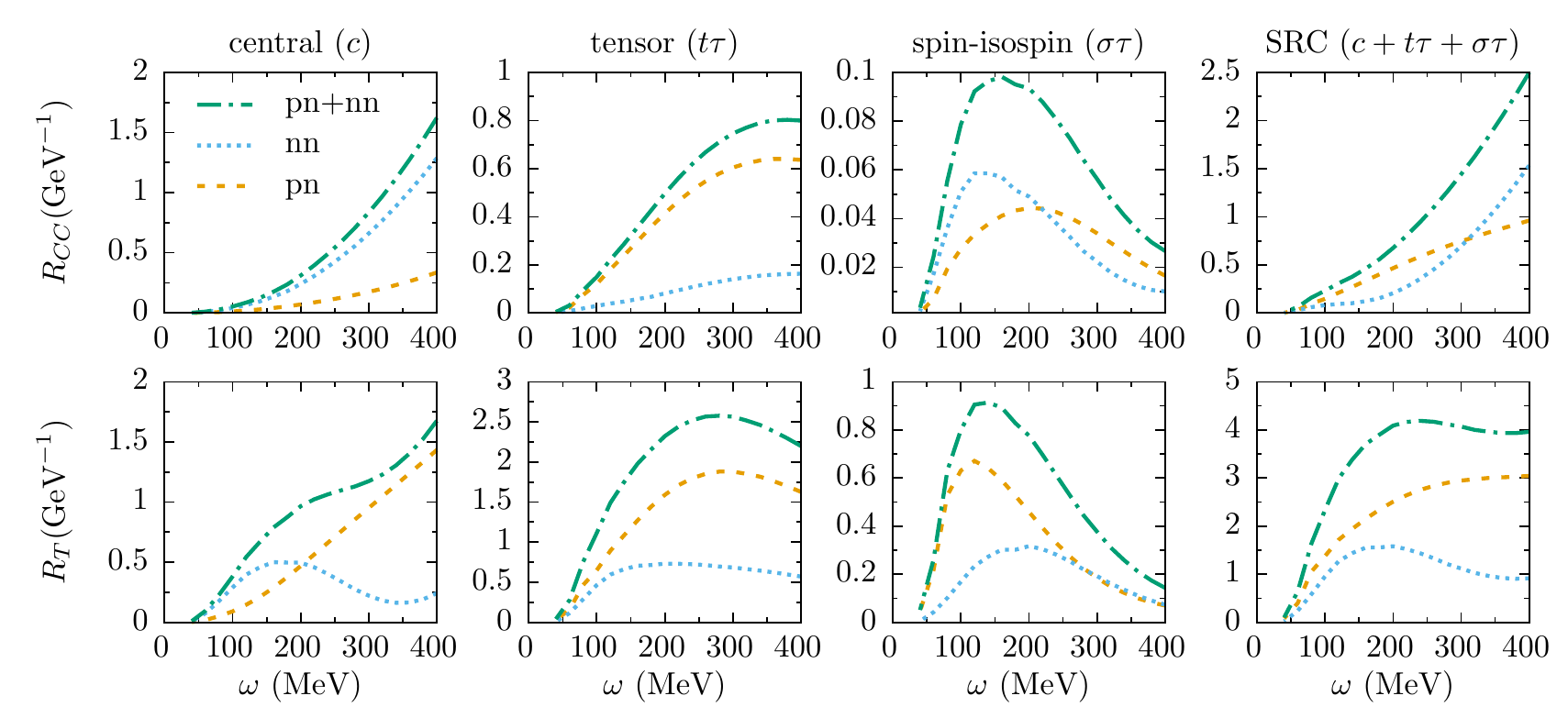}
   \caption{(Color online) Same as Fig. \ref{fig:neutincl4}. The contributions of the initial pn and nn pairs are shown for the three different SRC types.}
   \label{fig:neutincl5}
 \end{figure*}
 \begin{figure*}[t]
   \centering
   \includegraphics[width=0.95\textwidth,trim=0cm 1cm 0cm 0cm,clip=true]{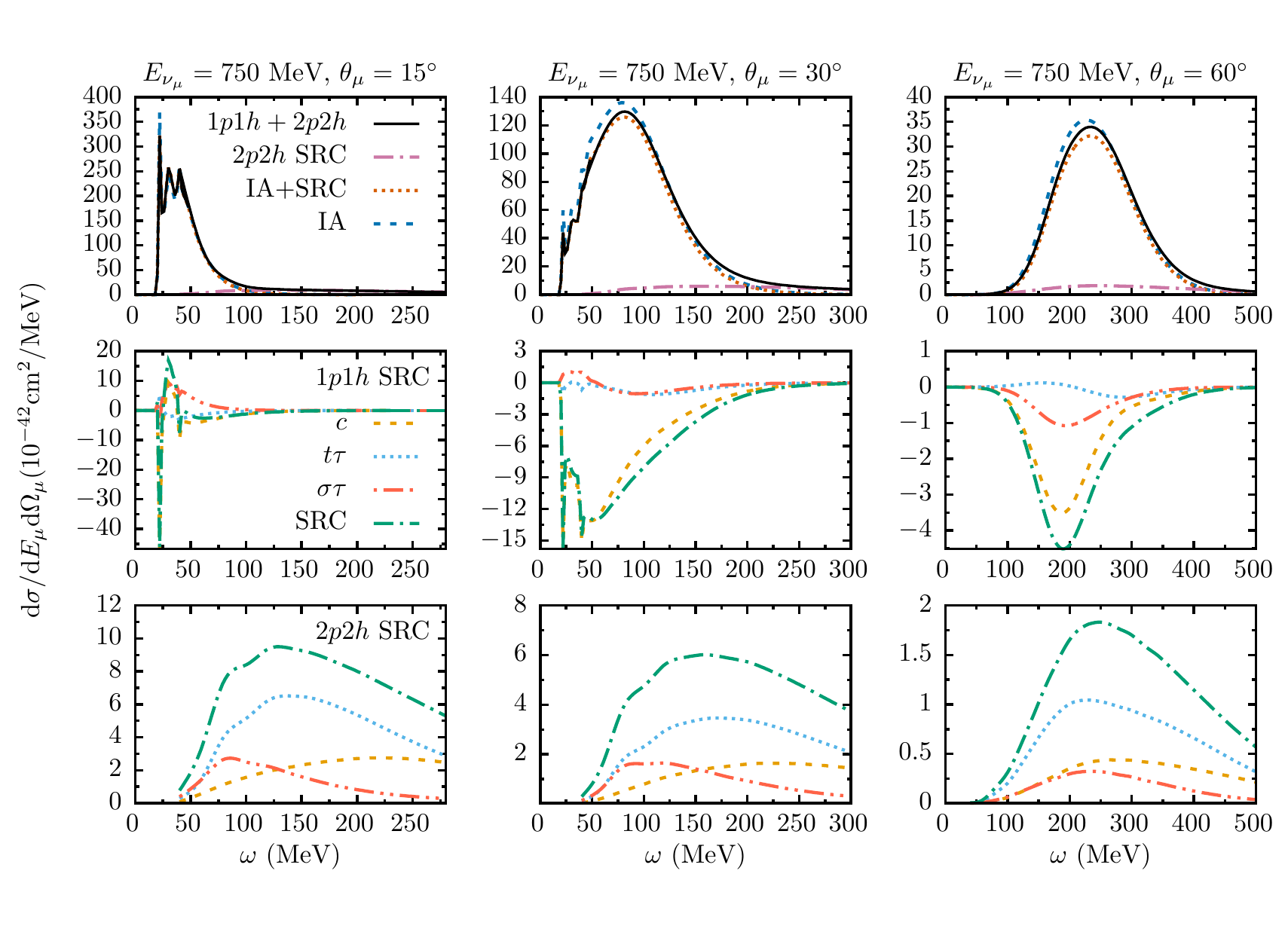}
   \caption{(Color online) The computed $\omega$ dependence of the $^{12}$C$(\nu_\mu,\mu^-)$ cross section for $E_{\nu_\mu}=750$ MeV and three different values for the lepton scattering angle $\theta_\mu$. The top panels show the combined 1p1h and 2p2h cross sections. The middle panels show the correction of the SRCs on the 1p1h cross section and the bottom panels show the 2p2h SRC part of the cross section.}
   \label{fig:neutincl6}
 \end{figure*}
 In the forthcoming, the results for quasielastic $^{12}$C$(\nu_\mu,\mu^-)$ cross sections with $1N$ and $2N$ knockout are presented. For neutrino interactions, the BBBA05 parameterization for the $Q^2$ dependence of the vector form factors is used \cite{Bradford:2006yz}. For the axial form factor $G_A$, the standard dipole form
 with $M_A = 1.03$ GeV is used. 

 The SRC induced 2p2h responses for CC neutrino interactions at fixed momentum transfer are shown in Figs.~\ref{fig:neutincl4} and \ref{fig:neutincl5}. 
 The Coulomb ($R_{CC}$) and transverse ($R_T$) response functions are presented to illustrate results for the time and space components of the nuclear current,  while maintaining a correspondence with electron scattering.
 In general, the $\omega$ dependence of the 2p2h responses does  not show a distinct peak as the 1p1h responses do, but continue to grow with increasing $\omega$. 
The reason of the broadening of the peak around $\omega=\frac{Q^2}{2m_N}$ 
for the 1p1h responses is the initial momentum of the interacting nucleon in the direction of the interacting neutrino, which lies within
 the interval $(-k_F,+k_F)$ with $k_F$ the Fermi momentum. For 2p2h 
 responses, the pairs initial momentum $P_{12}$ is the scaling variable. Momentum conservation poses no limits on the initial momenta of the separate particles, only on the momentum of the pair. The 2p2h responses of SRC pairs appear as a broad background ranging from the $2N$ knockout threshold to the maximum energy transfer, where $\omega = q$. Furthermore, the responses rise steadily with
 increasing $\omega$, which is the result of the growing phase-space. A similar, steadily increasing behavior of the 2p2h responses for electron scattering is seen in Refs.~\cite{VanOrden:1980tg,Alberico:1983zg,Gil:1997bm,Dekker:1994yc,DePace:2003xu} where the influence of MECs was studied.

 The separate contributions of the central ($c$), tensor ($t\tau$) and spin-isospin ($\sigma\tau$) correlations are shown in Fig.~\ref{fig:neutincl4}, for the vector and axial parts of the nuclear current. The tensor part yields the biggest contribution for small $\omega$ transfers, while the importance of the central part increases with $\omega$. This is directly related to the central and tensor correlation functions in
 momentum space. 
 In the axial part of the transverse response, the spin-isospin contribution is of similar size as the central and tensor correlations, while in the other channels (Coulomb and vector-transverse), the spin-isospin contribution is considerably smaller than the
 other two. This can be understood by looking at the operators of the spin-isospin correlation and the axial-transverse current. Both have a $\vec{\sigma} \cdot \vec{\tau}$ operator structure which strengthens the contribution. This dominance of the axial part over the vector part increases the importance of the spin-isospin correlations for neutrino compared to electron scattering.

 The strength attributed to the different initial pairs is shown in Fig.~\ref{fig:neutincl5}. The contributions are shown for the central, tensor and spin-isospin part for the SRCs. In the Coulomb response with central correlations, the contribution of initial nn pairs is roughly four times the contribution of the initial pn pairs. As the central correlation operator does not contain an isospin operator, it treats both protons and neutrons on an equal level. The factor four can be explained by noting that the $W^+$ boson only interacts with
 the neutrons in the initial pair, so that the nn matrix elements contain twice as many terms as the matrix elements for pn pairs. The tensor part is clearly dominated by pn pairs, as expected from its isospin structure.

  Finally, in Fig.~\ref{fig:neutincl6} we present the results for inclusive cross sections with $1N$ and $2N$ knockout for three different scattering angles. We have chosen an incoming neutrino energy of 750 MeV, which corresponds roughly with the peak of the MiniBooNE and T2K fluxes. The influence of SRCs on the 1p1h double differential cross section results in a small reduction, instead of the increase seen for electron scattering. The reason for this opposite
  behavior is related to the isospin part of the matrix elements and the different strength of the electric and magnetic form factors for electrons
  and neutrinos. Even when considering exclusively the vector part of the neutrino cross section, and treating the nucleons in the isospin formalism, the SRC correction for neutrinos has an opposite effect compared to electrons. The SRC correction is due to an interference between one-body and two-body matrix elements, where the sign of the isospin matrix element can result in either an increase or a decrease. 
 
 For the 2p2h part of the cross section, the contributions of the central, tensor and spin-isospin part of the correlations are shown separately. The tensor part is most important at small energy
 transfers but the relative importance of the central part grows for larger $\omega$, similar as seen in the responses separately. 
 The contribution of the spin-isospin correlations consists largely of the axial-transverse channel, as discussed earlier. 
 
 Comparing the position of the peak in the 1p1h and 2p2h channels, it is clear that the peak of the two-body channel occurs at higher $\omega$ than the QE-peak for small scattering angles. The difference decreases at higher scattering angles.
 For $\theta_\mu=60^\circ$ we remark that the reduction of the 1p1h channel and the contribution of the 2p2h channel have an opposite effect of similar size. The net effect of the short-range correlations on the inclusive signal is therefore rather small.

 \section{Summary}

 In this work, we have presented a model which accounts for SRCs in $\nu A$ scattering. Starting from HF nuclear wave functions, correlated nuclear wave functions are constructed. The correlations are taken into account by replacing the one-body nuclear current with an effective current. The expansion can be truncated at the two-body level owing to the
 local character of the SRCs.  

 The framework allows for the calculation of $1N$ and $2N$ knockout cross sections. The contribution of the $2N$ knockout channel to the inclusive cross section is presented. The integration over the solid angles of the two outgoing nucleons is performed analytically, the integration over their kinetic energy is performed numerically.

 The $^{12}$C$(e,e^\prime)$ results are compared with data. For neutrino scattering on $^{12}$C, the impact of the central, tensor and spin-isospin correlations were shown separately. The influences of the vector and axial-vector currents and the initial nucleon pair were studied as well. 
 
 The SRCs have a small influence on the $1N$ knockout channel and the SRC induced inclusive $2N$ knockout strength extends into the dip region of the double differential cross section. The SRCs similarly affect the vector and axial parts of the currents. 

 \section{Summary}

 In this work, we have presented a model which accounts for SRCs in $\nu A$ scattering. The technique was originally developed for exclusive $(e,e^\prime \textnormal{p} \textnormal{p})$ and semi-exclusive $(e,e^\prime \textnormal{p})$ scattering off $^{12}$C and $^{16}$O \cite{Ryckebusch:1997gn,Janssen:1999xy} and was compared with data \cite{Starink:2000aa,Ryckebusch:2003tu,Fissum:2004we,Iodice:2007mn}. Here we have extended the model to the weak CC interaction by
 including contributions from the axial vector current, which are absent in electromagnetic interactions.
 Starting from HF nuclear wave functions, correlated nuclear wave functions are constructed. The correlations are taken into account by replacing the one-body nuclear current with an effective current. The expansion can be truncated at the two-body level owing to the local character of SRCs. This formalism can be used for all target nuclei, for instance $^{40}$Ar which plays a major role in many recent and future neutrino experiments. 
 
 The framework allows for the calculation of $1N$ and $2N$ knockout cross sections. The contribution of the $2N$ knockout channel to the inclusive cross section is calculated by integrating over the phase space of the undetected nucleons. The integration over the solid angles of the two outgoing nucleons is performed analytically, the integration over their kinetic energy is performed numerically. The $^{12}$C$(e,e^\prime)$ results are compared with data. For neutrino scattering off $^{12}$C, the impact of the central, tensor and spin-isospin correlations
 were shown separately. The influences of the vector and axial-vector currents and the initial nucleon pair were studied as well. 
 
 The exclusive $2N$ knockout of SRC pairs shows a clear back-to-back signature which resembles the 'hammer events' seen by the ArgoNeuT collaboration \cite{Acciarri:2014gev}. The SRCs have a small influence on the $1N$ knockout channel and the SRC induced inclusive $2N$ knockout strength extends into the dip region of the double differential cross section. The $2N$ knockout strength from the vector and axial parts of the currents are of the same order of magnitude. For small $\omega$ values, the
 tensor correlations yield the biggest contribution while the importance of the central part increases with increasing $\omega$. This is a direct reflection of the
 properties of the central and tensor correlation functions in momentum space. The relative strength of the spin-isospin correlations for $\nu A$ scattering is larger compared to $eA$ scattering.
 
 It is normally assumed that, in the 2p2h channel, the majority of the cross section strength in the dip region comes from the MECs. Our results suggest an important role of the SRC induced $2N$ knockout. We conclude that SRCs and MECs should be considered consistently to fill the gap between theory and experiment. The study of these MECs for $\nu A$ processes is currently in progress.

 \appendix
 \section{Cross section}\label{app:cro}
 For $e A$ interactions, the kinematic factors in Eqs.~(\ref{eq:elecone}) and (\ref{eq:electwo}) are defined as
 \begin{align}
   v^e_L &= \frac{Q^4}{q^4}, \\
   v^e_T &= \frac{Q^2}{2q^2} + \tan^2\frac{\theta_{e'}}{2}, \\
   v^e_{TT} &= -\frac{Q^2}{2q^2}, \\
   v^e_{TL} &= -\frac{Q^2}{\sqrt{2}q^3}\left(E_e + E_{e'}\right) \tan^2\frac{\theta_{e'}}{2}.
 \end{align}
 For $\nu A$ interactions, the factors in Eqs.~(\ref{eq:neutone}) and (\ref{eq:neuttwo}) are given by (see e.g. Appendix A of \cite{Umino:1996cz})
 \begin{align}
   v_{CC} &= 1 + \zeta \cos\theta_\mu,   \\
   v_{CL} &= -\left( \frac{\omega}{q} \left(1 + \zeta \cos\theta_\mu \right) + \frac{m_\mu^2}{E_\mu q} \right), \\
   v_{LL} &= 1 + \zeta \cos\theta_\mu - \frac{2 E_{\nu_\mu} E_\mu}{q^2}\zeta^2 \sin^2\theta_\mu, \\
   v_T &= 1 - \zeta \cos\theta_\mu + \frac{E_{\nu_\mu} E_\mu}{q^2}\zeta^2 \sin^2\theta_\mu, \\
   v_{TT} &= -\frac{E_{\nu_\mu} E_\mu}{q^2}\zeta^2 \sin^2\theta_\mu, \\
   v_{TC} &= -\frac{\sin\theta_\mu}{\sqrt{2}q}\zeta\left(E_{\nu_\mu}+E_\mu\right), \\
   v_{TL} &=  \frac{\sin\theta_\mu}{\sqrt{2}q^2}\zeta\left(E_{\nu_\mu}^2-E_\mu^2 + m_\mu^2 \right), \\
   v_{T'} &= \frac{E_{\nu_\mu} + E_\mu}{q} \left( 1 - \zeta \cos\theta_\mu \right) - \frac{m_\mu^2}{E_\mu q}, \\
   v_{TC'} &= -\frac{\sin\theta_\mu}{\sqrt{2}} \zeta, \\
   v_{TL'} &= \frac{\omega}{q} \frac{\sin\theta_\mu}{\sqrt{2}} \zeta.
 \end{align}
 The nuclear response functions are identical for $e A$ and $\nu A$ interactions
 \begin{align} 
   W_{CC} &= \left| \mathcal{J}_0 \right|^2, \\
   W_{CL} &= 2\Re \left( \mathcal{J}_0 \mathcal{J}_3^\dagger \right), \\
   W_{LL} &= \left| \mathcal{J}_3 \right|^2, \\
   W_T &= \left| \mathcal{J}_{+1} \right|^2 + \left| \mathcal{J}_{-1} \right|^2, \\
   W_{TT} &= 2\Re \left( \mathcal{J}_{+1} \mathcal{J}_{-1}^\dagger \right), \\
   W_{TC} &= 2\Re \left[ \mathcal{J}_{0} \left( \mathcal{J}_{+1}^\dagger - \mathcal{J}_{-1}^\dagger \right) \right], \\
   W_{TL} &= 2\Re \left[ \mathcal{J}_{3} \left( \mathcal{J}_{+1}^\dagger - \mathcal{J}_{-1}^\dagger \right) \right], \\
   W_{T'} &= \left| \mathcal{J}_{+1} \right|^2 - \left| \mathcal{J}_{-1} \right|^2, \\
   W_{TC'} &= 2\Re \left[ \mathcal{J}_{0} \left( \mathcal{J}_{+1}^\dagger + \mathcal{J}_{-1}^\dagger \right) \right], \\
   W_{TL'} &= 2\Re \left[ \mathcal{J}_{3} \left( \mathcal{J}_{+1}^\dagger + \mathcal{J}_{-1}^\dagger \right) \right], \label{eq:responses}
 \end{align}
 with $\mathcal{J}_\lambda$ defined as in Eq.~(\ref{eq:currents}).

 \section{Matrix elements}\label{app:mat}
 In this appendix, we summarize the expressions for the 2p2h transition matrix elements with an effective two-body operator which accounts for SRCs. The standard expressions for the multipole operators and the nuclear currents are used (see e.g. Ref.~\cite{Walecka2004theoretical}) 
 \begin{align}
   \widehat{J}_0(q) &= +\sqrt{4\pi} \sum_{J \geq 0} i^J \widehat{J} \widehat{M}_{J0}^\textnormal{Coul} (q), \\
   \widehat{J}_3(q) &= -\sqrt{4\pi} \sum_{J \geq 0} i^J \widehat{J} \widehat{L}_{J0}^\textnormal{long} (q), \\
   \widehat{J}_{\pm1}(q) &= -\sqrt{2\pi} \sum_{J \geq 1} i^J \widehat{J} \left[\widehat{T}_{J\pm1}^\textnormal{elec}(q) \pm \widehat{T}_{J\pm1}^\textnormal{magn}(q)\right].
 \end{align}
 Here, the Coulomb operator is defined as
 \begin{align}
   \widehat{M}_{JM}^\textnormal{Coul}(q) = \int \textnormal{d} \vec{x} \left[ j_J(qx) Y_{JM}(\Omega_x) \right] \widehat{\rho}(\vec{x}).
 \end{align}
 Introducing the operator 
 \begin{align}
   \widehat{O}_{JM}^\kappa(q) = \sum_{M_1,M_2} \int \textnormal{d}\vec{x} \langle J+\kappa~M_1~1~M_2|J~M\rangle \nonumber \\
   \times \left[ j_{J+\kappa}(qx) Y_{J+\kappa M_1}(\Omega_x) \right] \widehat{J}_{M_2}(\vec{x}),
 \end{align}
 the longitudinal, electric and magnetic transition operators are written as
 \begin{align}
   \widehat{L}_{JM}^\textnormal{long}(q) &= i \sum_{\kappa=\pm1} \frac{ \sqrt{ J+\delta_{\kappa,+1}}}{\widehat{J}} \widehat{O}_{JM}^\kappa(q), \\
   \widehat{T}_{JM}^\textnormal{elec}(q) &= i \sum_{\kappa=\pm1} (-1)^{\delta_{\kappa,+1}} \frac{ \sqrt{ J+\delta_{\kappa,-1}}}{\widehat{J}} \widehat{O}_{JM}^\kappa(q), \\
   \widehat{T}_{JM}^\textnormal{magn}(q) &= \widehat{O}_{JM}^{\kappa=0}(q).
 \end{align}

 Hence, matrix elements of the operator $\widehat{O}_{JM}^\kappa$ suffice to determine the strengths of the longitudinal, electric and magnetic transition operators. In the matrix elements, we used the shorthand notation $a \equiv (n_a,l_a,1/2,j_a)$.
 The operators $\widehat{\rho}(\vec{x})$ and $\widehat{J}(\vec{x})$ in the definitions of $\widehat{M}$ and $\widehat{O}$ are the time and space component of the nuclear current operator in coordinate space. The matrix elements accounting for the vector parts of the nuclear current, $\widehat{J}_\mu^V(\vec{x})$, are given in Refs.~\cite{Ryckebusch:1997gn} and \cite{Janssen:1999xy} for central and spin-dependent correlations in electron scattering respectively. They can be translated
 into neutrino
 interactions after a rotation in isospin space.
 The matrix elements for the axial parts,
 $\widehat{J}_\mu^A(\vec{x})$, are given below.  
 We will first consider the matrix elements for central correlations and afterwards those for tensor and spin-isospin correlations. The expressions below are given for CC neutrino interactions. The $\tau_\pm$ operator is responsible for the flavor change induced by the $W^\pm$ boson. 
 \begin{widetext}
 \subsection{Central correlations}
 The partial-wave components of the central correlation function are obtained via
 \begin{align}
    \chi^c(l,r_i,r_j) &= \frac{2l+1}{2} \int_{-1}^{+1} \textnormal{d} \cos \theta \, P_l(\cos \theta) g_c\left(\sqrt{r_i^2+r_j^2-2r_ir_j \cos \theta}\right),
 \end{align}
with $P_l(x)$ the Legendre polynomial of order $l$. The axial 2p2h matrix elements arising from the coupling of a one-body current in the IA to a central-correlated pair are given by
   \begin{align}
     \langle ab ; J_1 || \widehat{M}_J^\textnormal{Coul} \left[ \widehat{\rho}_{A}^{[1],c}(i,j) \right] || cd ; J_2 \rangle &= - \frac{G_A}{m_N i} \sqrt{\pi} \sum_{l,L} \frac{\widehat{J}_1 \widehat{J}_2 \widehat{L}}{\widehat{l}} \langle L~0~l~0 | J~0 \rangle \int \mathrm{d} r_i \int \mathrm{d} r_j ~\chi^c(l,r_i,r_j) \nonumber \\
     \times \Biggl( & \langle a || \tau_\pm || c \rangle \langle a || j_J (q r_i) Y_L (\Omega_i) || c \rangle_{r_i} \langle b || Y_l (\Omega_j) || d \rangle_{r_j} 
     \begin{Bmatrix} 
       j_a & j_b & J_1 \\ j_c & j_d & J_2 \\ L & l & J 
     \end{Bmatrix} \nonumber \\
     + & \langle b || \tau_\pm || d \rangle \langle a ||  Y_l (\Omega_i) || c \rangle_{r_i} \langle b || j_J (q r_j) Y_L (\Omega_j) || d \rangle_{r_j} 
     \begin{Bmatrix} 
       j_a & j_b & J_1 \\ j_c & j_d & J_2 \\ l  & L & J 
     \end{Bmatrix} \Biggr),
   \end{align}
   \begin{align}
     \langle ab ; J_1 || \widehat{O}_J^{\kappa} \left[ \widehat{J}_{A}^{[1],c}(i,j) \right] || cd ; J_2 \rangle &= G_A \sqrt{4 \pi} \sum_{l,L,J_x} \frac{\widehat{L} \widehat{J}_x \widehat{J}_1 \widehat{J}_2 \widehat{J}}{\widehat{l}} 
     \begin{Bmatrix}
       L & l & J+\kappa \\ J & 1 & J_x 
     \end{Bmatrix}
     \int \mathrm{d} r_i \int \mathrm{d} r_j ~\chi^c(l,r_i,r_j) \nonumber \\
     \times \Biggl( (-1)^{(J_x+L)} & \langle a || \tau_\pm || c \rangle \langle a || j_{J+\kappa} (q r_i) \left[ Y_L (\Omega_i) \otimes \vec{\sigma}_i \right]_{J_x} || c \rangle_{r_i} \langle b || Y_l (\Omega_j) || d \rangle_{r_j} 
     \begin{Bmatrix} 
       j_a & j_b & J_1 \\ j_c & j_d & J_2 \\ J_x & l & J 
     \end{Bmatrix} \nonumber \\
     + (-1)^{(L+l +J)} & \langle b || \tau_\pm || d \rangle \langle a ||  Y_l (\Omega_i) || c \rangle_{r_i} \langle b || j_{J+\kappa} (q r_j) \left[ Y_L (\Omega_j) \otimes \vec{\sigma}_j \right]_{J_x} || d \rangle_{r_j} 
     \begin{Bmatrix} 
       j_a & j_b & J_1 \\ j_c & j_d & J_2 \\ l  & J_x & J 
     \end{Bmatrix} \Biggr).
   \end{align}
   The radial transition densities $\langle a || \widehat{\mathcal{O}}_J || b \rangle_r$ are defined so that they are related to the full matrix elements as $\langle a || \widehat{\mathcal{O}}_J || b \rangle \equiv \int \textnormal{d} r \langle a || \widehat{\mathcal{O}}_J || b \rangle_r$.

 \subsection{Tensor correlations}
 The partial-wave components of the tensor correlation function are defined as
 \begin{align}
   \chi^{t\tau}(l_1,l_2,r_i,r_j) &= \int \textnormal{d} q \int \textnormal{d} r q^2 r^2 j_2(qr) j_{l_1}(qr_i) j_{l_2}(qr_j) f_{t\tau} \left( r \right).
 \end{align}
 The axial transition matrix elements accounting for the coupling of a one-body current to a tensor-correlated pair are given by
   \begin{align}
     \langle ab & ; J_1 || \widehat{M}_J^\textnormal{Coul} \left[ \widehat{\rho}_{A}^{[1],t\tau}(i,j) \right] || cd ; J_2 \rangle = G_A \frac{2\sqrt{6}}{\sqrt{\pi} m_N i}  \sum_{l_1,l_2} \sum_{J_3,J_4} \sum_L \int \mathrm{d} r_i \int \mathrm{d} r_j~\widehat{l}_1 \widehat{l}_2 \widehat{L} \widehat{J} \widehat{J}_1 \widehat{J}_2 \widehat{J}_3 \widehat{J}_4 \nonumber \\
     & \times \langle l_1~0~l_2~0 | 2~0 \rangle 
     \begin{Bmatrix}
       1 & 1 & 2 \\ l_1 & l_2 & J_3
     \end{Bmatrix}
     i^{l_1 + l_2}~\chi^{t \tau}(l_1,l_2,r_i,r_j) \nonumber \\
     \times \Biggl( & \langle ab || \tau_\pm(1) \left(\vec{\tau}_1 \cdot \vec{\tau}_2 \right) || cd \rangle
     \begin{pmatrix}
       L & J & l_1 \\ 0 & 0 & 0
     \end{pmatrix}
     \begin{Bmatrix}
       L & J & l_1 \\ J_3 & 1 & J_4
     \end{Bmatrix}
     \begin{Bmatrix} 
       j_a & j_b & J_1 \\ j_c & j_d & J_2 \\ J_4 & J_3 & J 
     \end{Bmatrix} \widehat{l}_1 (-1)^{J+1} \nonumber \\
     & \qquad \qquad \qquad \times \langle a || j_J (q r_i) \left[ Y_L (\Omega_i) \vec{\sigma}_i \left( \overrightarrow{\nabla}_i - \overleftarrow{\nabla}_i \right) \otimes \vec{\sigma}_i \right]_{J_4} || c \rangle_{r_i} \langle b || \left[ Y_{l_2} (\Omega_j) \otimes \vec{\sigma}_j \right]_{J_3} || d \rangle_{r_j} \nonumber \\
     + & \langle ab || \tau_\pm(2) \left(\vec{\tau}_1 \cdot \vec{\tau}_2 \right)|| cd \rangle
     \begin{pmatrix}
       L & J & l_2 \\ 0 & 0 & 0
     \end{pmatrix}
     \begin{Bmatrix}
       L & J & l_2 \\ J_3 & 1 & J_4
     \end{Bmatrix}
     \begin{Bmatrix} 
       j_a & j_b & J_1 \\ j_c & j_d & J_2 \\ J_3 & J_4 & J 
     \end{Bmatrix}\widehat{l}_2 (-1)^{J_3+J_4+1} \nonumber \\
     & \qquad \qquad \qquad \times \langle a || \left[ Y_{l_1} (\Omega_i) \otimes \vec{\sigma}_j \right]_{J_3} || c \rangle_{r_i} \langle b || j_J (q r_j) \left[ Y_L (\Omega_j) \vec{\sigma}_j \left( \overrightarrow{\nabla}_j - \overleftarrow{\nabla}_j \right) \otimes \vec{\sigma}_j \right]_{J_4} || d \rangle_{r_j}  \Biggr),
   \end{align}
   \begin{align}
     \langle ab & ; J_1 || \widehat{O}_J^{\kappa} \left[ \widehat{J}_A^{[1],t\tau}(i,j) \right] || cd ; J_2 \rangle = G_A \frac{12}{\sqrt{\pi}} \sum_{l_1,l_2} \sum_{J_3,J_4} \sum_{J_5,L} \int \mathrm{d} r_i \int \mathrm{d} r_j~\widehat{l}_1 \widehat{l}_2 \widehat{L} \widehat{J} \widehat{J}_1 \widehat{J}_2 \widehat{J}_3 \widehat{J}_4 \
     \left( \widehat{J}_5 \right)^2 \nonumber \\
     & \qquad \qquad \qquad \times \widehat{J+\kappa}~\langle l_1~0~l_2~0 | 2~0 \rangle 
     \begin{Bmatrix}
       1 & 1 & 2 \\ l_2 & l_1 & J_3
     \end{Bmatrix}
     i^{l_1 + l_2-1}~\chi^{t \tau}(l_1,l_2,r_i,r_j)\nonumber \\
     \times \Biggl( (-1)^{J} \widehat{l}_1 \widehat{j}_a \widehat{j}_c  & \langle ab || \tau_\pm(1) \left(\vec{\tau}_1 \cdot \vec{\tau}_2 \right) || cd \rangle
     \begin{pmatrix}
       L & l_1 & J+\kappa \\ 0 & 0 & 0
     \end{pmatrix}
     \begin{Bmatrix}
       1 & J & J+\kappa \\ l_1 & L & J_4
     \end{Bmatrix}
     \begin{Bmatrix}
       1 & J_3 & l_1 \\ J & J_4 & J_5
     \end{Bmatrix}
     \begin{Bmatrix} 
       j_a & j_b & J_1 \\ j_c & j_d & J_2 \\ J_5 & J_3 & J 
     \end{Bmatrix}
     \begin{Bmatrix} 
       l_a & 1/2 & j_a \\ l_c & 1/2 & j_c \\ J_4 & 1 & J_5 
     \end{Bmatrix} \nonumber \\
     \times &  \langle n_a l_a || j_{J+\kappa} (q r_i) \left[ Y_L (\Omega_i) \otimes \vec{\sigma}_i \right]_{J_4} || n_c l_c \rangle_{r_i} \langle b || \left[ Y_{l_2} (\Omega_j) \otimes \vec{\sigma}_j \right]_{J_3} || d \rangle_{r_j} \nonumber \\
     + (-1)^{J_3+J_5} \widehat{l}_2 \widehat{j}_b \widehat{j}_d  & \langle ab || \tau_\pm(2)  \left(\vec{\tau}_1 \cdot \vec{\tau}_2 \right)|| cd \rangle
     \begin{pmatrix}
       L & l_2 & J+\kappa \\ 0 & 0 & 0
     \end{pmatrix}
     \begin{Bmatrix}
       1 & J & J+\kappa \\ l_2 & L & J_4
     \end{Bmatrix}
     \begin{Bmatrix}
       1 & J_3 & l_2 \\ J & J_4 & J_5
     \end{Bmatrix}
     \begin{Bmatrix} 
       j_a & j_b & J_1 \\ j_c & j_d & J_2 \\ J_3 & J_5 & J 
     \end{Bmatrix}
     \begin{Bmatrix} 
       l_b & 1/2 & j_b \\ l_d & 1/2 & j_d \\ J_4 & 1 & J_5 
     \end{Bmatrix} \nonumber \\
     \times & \langle a || \left[ Y_{l_2} (\Omega_i) \otimes \vec{\sigma}_i \right]_{J_3} || c \rangle_{r_i}  \langle n_b l_b || j_{J+\kappa} (q r_j) \left[ Y_L (\Omega_j) \otimes \vec{\sigma}_j \right]_{J_4} || n_d l_d \rangle_{r_j} \Biggr).
   \end{align}
 The operators $\overrightarrow{\nabla}$ and $\overleftarrow{\nabla}$ refer to the gradient operators acting to the right and left respectively.

 \subsection{Spin-isospin correlations}
 The partial-wave components of the spin-isospin correlation function are defined as
 \begin{align}
   \chi^{\sigma \tau}(l,r_i,r_j) &= \int_{-1}^{+1} \textnormal{d} \cos \theta \, P_l(\cos \theta) f_{\sigma \tau} \left( \sqrt{r_i^2+r_j^2-2r_ir_j \cos \theta} \right).
 \end{align}
 The axial matrix elements describing the effective coupling of a virtual boson to a spin-isospin correlated nucleon pair are given by

   \begin{align}
     \langle ab & ; J_1 \parallel  \widehat{M}_{J}^{\rm{Coul}} \left[ \widehat{\rho}_{A}^{[1],\sigma\tau}(i,j) \right] \parallel cd ; J_2 \rangle = G_A \frac{\sqrt{\pi}}{m_N i} \sum_{l,L} \sum_{J_3,J_4} \int \textnormal{d} r_i \int \textnormal{d} r_j   \frac{\widehat{L} \widehat{J}_1 \widehat{J}_2 \widehat{J}_3 \widehat{J}_4}{\widehat{l}} \nonumber \\
& \times \langle l~0~L~0 | J~0 \rangle
\begin{Bmatrix}
  J_3 & L & 1 \\
  l & J_4 & J 
\end{Bmatrix}
\ \chi^{\sigma \tau}\left( l, r_i,  r_j \right) 
  \nonumber \\ 
 \times \Biggl(  & \langle ab |\tau_\pm(1) \left(\vec{\tau}_1 \cdot \vec{\tau}_2 \right) | cd \rangle\
 \begin{Bmatrix}
   j_a & j_b & J_1 \\
   j_c & j_d & J_2 \\
   J_3 & J_4 & J
 \end{Bmatrix}
 \left( -1 \right)^{l+J_4}\nonumber \\ 
 & \qquad \qquad \qquad \times \langle  a \parallel j_J \left( qr_1 \right)  \left[ Y_L  \left( \Omega_1 \right)\vec{\sigma}_1 \left( \overrightarrow{{\nabla}}_1 - \overleftarrow{{\nabla}}_1 \right) \otimes \vec{ \sigma}_1 \right]_{J_3}  \parallel c \rangle_{r_i} \langle b \parallel \left[ Y_{l} \left( \Omega_2 \right) \otimes \vec{ \sigma}_2 \right]_{J_4} \parallel d \rangle_{r_j} \nonumber \\
 +  & \langle ab |\tau_\pm(2)\left(\vec{\tau}_1 \cdot \vec{\tau}_2 \right) | cd \rangle 
 \begin{Bmatrix}
   j_a & j_b & J_1 \\
   j_c & j_d & J_2 \\
   J_4 & J_3 & J
 \end{Bmatrix}
\left( -1 \right)^{l+J+J_3}\nonumber \\
& \qquad \qquad \qquad \times \langle  a \parallel \left[ Y_{l} \left( \Omega_1 \right) \otimes \vec{  \sigma}_1 \right]_{J_4}  \parallel c \rangle_{r_i} \langle b \parallel j_J \left( qr_2 \right)  \left[ Y_L \left( \Omega_2 \right) \vec{\sigma}_2 \left( \overrightarrow{{\nabla}}_2 - \overleftarrow{{\nabla}}_2 \right)\otimes \vec{ \sigma}_2 \right]_{J_3}
\parallel d \rangle_{r_j}  \Biggr),\label{eq:spicoul}
   \end{align}
  \begin{align}
    \langle ab & ; J_1 \parallel \widehat{O}_J^{\kappa} \left[ \widehat{J}_A^{[1],\sigma\tau}(i,j)\right] \parallel cd ; J_1 \rangle = G_A \sqrt{24 \pi} \sum_{l,L} \sum_{J_4,J_5} \sum_{J_6} \sum_j \int \textnormal{d} r_1 \int \textnormal{d} r_2 \frac{ \widehat{L} \widehat{J}  \widehat{J+\kappa} \widehat{J}_1 \widehat{J}_2 \left( \widehat{J}_4 \right)^2 \widehat{J}_5 \widehat{J}_6 }{\widehat{l}} \nonumber \\
    & \times 
    \begin{pmatrix}
      J+\kappa & L & l \\
      0 & 0 & 0 
    \end{pmatrix}
    \begin{Bmatrix}
      J & 1 & J+\kappa \\
      L & l & J_6
    \end{Bmatrix}
    \begin{Bmatrix} 
      J_6 & l & J \\
      J_5 & J_4 & 1
    \end{Bmatrix}
    \ \chi^{\sigma \tau}(l, r_i, r_j) \nonumber \\
    \times \Biggl( & \langle ab |\tau_\pm(1)\left(\vec{\tau}_1 \cdot \vec{\tau}_2 \right) | cd \rangle 
    \begin{Bmatrix}
      J_6 & 1 & J_4 \\
      j_c & j_a & j
    \end{Bmatrix}
    \begin{Bmatrix}
      1/2 & j & l_c \\
      j_c & 1/2 & 1
    \end{Bmatrix}
 \begin{Bmatrix}
   j_a & j_b & J_1 \\
   j_c & j_d & J_2 \\
   J_4 & J_5 & J
 \end{Bmatrix}
    \widehat{j} \widehat{j}_c (-1)^{L+J_4+J_5+j_a+j_c+j+l_c+3/2} \nonumber \\
    & \qquad \qquad \qquad \times \langle a \parallel j_{J+\kappa}(qr_1) \left[ Y_L(\Omega_1) \otimes \vec{\sigma}_1 \right]_{J_6} \parallel n_c l_c \tfrac{1}{2} j \rangle_{r_i} \ \langle b \parallel \left[ Y_l(\Omega_2) \otimes \vec{\sigma}_2 \right]_{J_5} \parallel d \rangle_{r_j} \nonumber \\
    \times & \langle ab |\tau_\pm(2)\left(\vec{\tau}_1 \cdot \vec{\tau}_2 \right) | cd \rangle  
    \begin{Bmatrix}
      J_6 & 1 & J_4 \\
      j_d & j_b & j
    \end{Bmatrix}
    \begin{Bmatrix}
      1/2 & j & l_d \\
      j_d & 1/2 & 1
    \end{Bmatrix}
 \begin{Bmatrix}
   j_a & j_b & J_1 \\
   j_c & j_d & J_2 \\
   J_5 & J_4 & J
 \end{Bmatrix}
    \widehat{j} \widehat{j}_d (-1)^{L+J+j_b+j_d+j+l_d+3/2} \nonumber \\
    & \qquad \qquad \qquad \times \langle a \parallel \left[ Y_l(\Omega_1) \otimes \vec{\sigma}_1 \right]_{J_5} \parallel c \rangle_{r_i} \ \langle b \parallel j_{J+\kappa}(qr_2)\left[ Y_L(\Omega_2) \otimes \vec{\sigma}_2 \right]_{J_6} \parallel n_d l_d \tfrac{1}{2} j \rangle_{r_j} \Biggr).
  \end{align}

 \end{widetext}

 \begin{acknowledgments}
   This work was supported by the Interuniversity Attraction Poles Programme P7/12 initiated by the Belgian Science Policy Office and the Research Foundation Flanders (FWO-Flanders). The computational resources (Stevin Supercomputer Infrastructure) and services used in this work were provided by Ghent University, the Hercules Foundation and the Flemish Government.
 \end{acknowledgments}

 \bibliography{biblio}

 \end{document}